\documentclass[usenatbib]{mn2e}
\usepackage{times}
\usepackage{graphicx}
\usepackage{natbib}
\usepackage{subfigure}
\usepackage{dcolumn}
\usepackage{longtable}
\usepackage{lscape}
\usepackage[usenames,dvipsnames]{xcolor}
\usepackage{xspace}
\usepackage{aas_macros}
\usepackage{xargs}
\usepackage{url}

\newcommand{\galfit}{{\scshape galfit}\xspace}
\newcommand{\galfitthree}{{\scshape galfit3}\xspace}
\newcommand{\galfitm}{{\scshape galfitm}\xspace}

\newcommand{\megamorph}{{MegaMorph}\xspace}
\newcommand{\sersic}{S\'ersic\xspace}

\newcommand{\montage}{{\scshape Montage}\xspace}
\newcommand{\ferengi}{{\scshape ferengi}\xspace}

\newcommand{\fitb}{\texttt{SM}\xspace}
\newcommand{\fitc}{\texttt{MM}\xspace}
\newcommand{\re}{r_{\rm e}}
\newcommand{\PA}{P\!A}
\newcommandx{\N}[2][1= ,2= ]{$\mathcal{N}^{#1}_{#2}$\xspace}
\newcommandx{\R}[2][1= ,2= ]{$\mathcal{R}^{#1}_{#2}$\xspace}

\newcolumntype{d}{D{.}{.}{-1}}

\title[MegaMorph]{MegaMorph -- multi-wavelength measurement of galaxy structure: physically meaningful bulge-disc decomposition of galaxies near and far}
\author[Vika et al.]
{Marina~Vika,$^{1}$  Steven~P.~Bamford,$^{2}$  Boris~H{\"a}u{\ss}ler,$^{3,4}$ Alex~L.~Rojas$^{1}$
\smallskip\\
$^1$Carnegie Mellon University in Qatar, Education City, PO Box 24866, Doha, Qatar \\
$^2$School of Physics and Astronomy, The University of Nottingham, University Park, Nottingham, NG7 2RD, UK \\
$^3$Department of Physics, University of Oxford, Denys Wilkinson Building, Keble Road, Oxford, OX1 3RH, UK \\
$^4$University of Hertfordshire, Hatfield, Hertfordshire, AL10 9AB, UK
}

\begin{document}

\date{Accepted ... Received ...; in original form ...}

\pagerange{\pageref{firstpage}--\pageref{lastpage}} \pubyear{2012}

\maketitle

\label{firstpage}

\begin{abstract}

Bulge-disc decomposition is a valuable tool for understanding galaxies.  However, achieving robust measurements of component properties is difficult, even with high quality imaging, and it becomes even more so with the imaging typical of large surveys.

In this paper we consider the advantages of a new, multi-band approach to galaxy fitting.  We perform automated bulge-disc decompositions for 163 nearby galaxies, by simultaneously fitting multiple images taken in five photometric filters.  We show that we are able to recover structural measurements that agree well with various other works, and confirm a number of key results.  We additionally use our results to illustrate the link between total \sersic index and bulge-disc structure, and demonstrate that the visually classification of lenticular galaxies is strongly dependent on the inclination of their disc component.

By simulating the same set of galaxies as they would appear if observed at a range of redshifts, we are able to study the behaviour of bulge-disc decompositions as data quality diminishes.  We examine how our multi-band fits perform, and compare to the results of more conventional, single-band methods.  Multi-band fitting improves the measurement of all parameters, but particularly the bulge-to total flux ratio and component colours.  We therefore encourage the use of this approach with future surveys.

\end{abstract} 


\section{Introduction}
\label{sec:intro}

The spatial distribution of  light within a galaxy is a key observable, with which we can constrain models of galaxy formation and evolution.  The typical sizes, surface-brightness profiles and ellipticities of galaxies have been essential in determining the main physical mechanisms at work in producing the galaxy population (e.g., \citealt{tex:SN81}).  We have grown to understand that these properties are the result of multiple competing processes, including rapid collapse, ongoing gas accretion, disk instabilities, and the merging of existing stellar systems (e.g., \citealt{tex:B10}).  As a consequence, galaxies are often separable, at least to a degree, into components with distinct spatial structure, kinematics and stellar populations.

Observations often integrate over these components, e.g. aperture photometry, to estimate overall properties for each galaxy.  With such quantities, one can gain a general picture of the merger and star-formation history of a given galaxy.  Typically, however, a variety of histories can produce similar integrated properties.  Considering the properties of a galaxy's components separately enables a much more detailed account of its lifetime to be constructed.


The simplest approach to separating the properties of the main galaxy structures is bulge-disc decomposition.  This can be applied to imaging data alone, and hence to the largest samples of galaxies available.
Although conceptually simple, bulge-disc decomposition remains a challenging task, due to the variety of structures that galaxies display, not to mention the usual observational limitations of resolution and signal-to-noise.

A common method is to study and model the one dimensional (1D) light profile, along the major or minor axis of the galaxy, or azimuthally averaged.  These 1D profiles are usually obtained by fitting a set of ellipses to the isophotes in the (2D) image. 
However, 1D representations of the radial surface-brightness distribution suffer from strong systematic uncertainties since they neglect the differing intrinsic shapes of the disk and bulge components.

A solution to this problem is two-dimensional (2D) decomposition \citep{tex:BF95}, which utilises all the spatial information in the images (for more details about standard 1D and 2D methods see \citealt{tex:PH10}).  On the other hand, fitting in 2D is usually more sensitive to features, such as bars and spiral arms, which are difficult to model.
The usual procedure for 2D bulge-disc decomposition is to fit a parametric model to the image, accounting for the point spread function (PSF), pixelisation and noise properties of the image.  The projected surface-brightness profile of each component is typically modelled using an analytic function, the most common choice being the \sersic profile \citep{tex:S68}.


Separating galaxy components is supposedly easier for small samples of nearby galaxies where a more interactive fitting process can be applied. 
Multiple studies have applied 2D decomposition to examine the correlations between bulge and disc properties at optical to infrared wavelengths (e.g., \citealt{tex:Nv07,tex:MA08,tex:BW09,tex:TW11}), to study the coevolution of supermassive black holes and their host galaxy (e.g., \citealt{tex:KH08,tex:VD12}), to investigate the evolution of structure over cosmic time (e.g., \citealt{tex:BD12,tex:BD14}) and environment (\citealt{tex:HS10,tex:HL14}), to study the structural properties of isolated late type galaxies (e.g., \citealt{tex:DS08}), and to measure quasar host galaxy parameters (e.g., \citealt{tex:MD00}). 

Some studies go a step further and attempt to decompose a third component, usually a bar (e.g., \citealt{tex:LS05,tex:G09,tex:WJ09,tex:G11b}).
In addition to providing measurements of bar properties for study, including a potential bar in the model helps to avoid any such feature from contaminating measurements of the bulge and disc.

A significant issue lies in identifying which components are present, and hence which model parameters are to be trusted.  This amounts to choosing the appropriate complexity of model for a given galaxy.  Fitting a more complex model usually results in a significantly improved goodness-of-fit statistic (e.g., chi-squared), irrespective of whether or not the model parameters are physically meaningful.  This problem is complicated by the presence of galaxy features that are not included in the model, such as cores, non-elliptical and twisted isophotes, dust lanes, etc.  Many studies ultimately resort to selecting the most appropriate model by visual inspection of the original images and their fit residuals.  

Elliptical galaxies are usually regarded to be one-component systems, and hence they are usually chosen to be modelled by a single \sersic profile.  However, it is far from clear whether this is physically the best way to describe these systems.  Taking a different approach, \citet{tex:HH13} fitted three components to each member of a sample of elliptical galaxies, finding that these galaxies can be well described by the combination of three \sersic profiles, each with low \sersic index but different effective radii. \citeauthor{tex:HH13} argue that these components have physically meaningful interpretations. 
The intermediate-size component is the original, built from early collapse and major mergers.  The largest component is comprised of stars accreted in more recent minor mergers.   
Finally, the most compact component is attributed to central star formation following the dissipative accretion of gas brought in by some of those recent minor mergers.

For large samples of galaxies, more automated approaches to deciding how many components a galaxy comprises are essential.   For example,
\citet{tex:AD06} employed a logical filter to decide whether the results of fitting a bulge-disc model were physically plausible, or whether their single-\sersic fit should be preferred.  They showed that the routine structural decomposition is an important for understanding the bimodality of galactic properties. 
\citet{tex:SM11} have created the largest catalogue of multi-component galaxy structure to date.  They fit one million galaxies with three different models, and used F-tests with a calibrated probability threshold to choose the best model for each galaxy. \citet{tex:LG12} expanded the model options five, selecting between them using a logical filter. These studies have provided the first complete estimates of the bulge and disc properties for the local Universe.  

  
To date, most studies have measured structural properties of a galaxy using only one image, in a single waveband. However, modern surveys provide images of the same galaxies in many different bands.  In some cases, models are fit to each band independently.  This does not produce reliable colours, however, so more often an initial model is fit to one preferred band, then the structural parameters are fixed during fits to the other bands.  \citet{tex:SM11}  (following \citet{tex:SW02}) take a more consistent approach by fitting their models to images in two bands simultaneously, while \citet{tex:MS13} use a hybrid procedure to produce bulge and disc colours in five optical bands.


Until recently, no method was available that could fit models to an arbitrary number of images at different wavelengths.  Driven by a determination to make more effective use of the multi-wavelength imaging available from modern surveys, the \megamorph project (\citealt{tex:HB13, tex:VB13b} and Bamford, in prep.) developed and tested a new version of two-dimensional photometric analysis which constrains a single, wavelength-dependent model using multiple images simultaneously.

This paper is one of a series that investigates the benefits of this multi-wavelength approach to measuring galaxy structural properties. In Bamford et al. (in prep.) we present this new tool in detail, describing the new features and demonstrating its use through some specific examples. In \citet[hereafter V13]{tex:VB13b}  we test our new method by fitting single-\sersic models to original and artificially-redshifted image of 163 nearby galaxies.
In \citet{tex:HB13}  we demonstrate our approach on a large dataset from the GAMA \citep{tex:DN09,tex:HK10} survey, automating both the preparation of the data and the fitting process itself. The resulting measurements -- in particular the variation of structural parameters with wavelength -- are studied further in \citet{tex:VB14}.
The objective of the present paper, is to investigate the ability of \galfitm to perform bulge-disc decomposition on galaxy images with a wide range of resolution and signal-to-noise.
This is achieved by analysing the same sample as V13: large, nearby galaxies in the Sloan Digital Sky Survey (SDSS; \citealt{tex:AA09}), with both original images and versions that have been convolved and resampled in order to simulate the galaxies' appearance at a range of redshifts. 
A complementary analysis of multi-band bulge-disc decomposition, using the same GAMA sample as \citet{tex:HB13}, will be presented in a forthcoming paper (Haeussler et al., in prep.).

This paper is structured as follows: in Section 2 we present our data set, give a brief description of \galfitm, and then explain how we fit our sample and identify reliable components.  
In Section 3 we present the distributions of structural parameters obtained from the original SDSS imaging, and examine the stability of these distributions with respect to the effects of distance. 
In Section 4 we present correlation between structural parameters and a  way of separating elliptical from lenticular galaxies in our sample.
We provide a summary in Section 5.

\section{Data}

\subsection{Sample selection and imaging} 

\begin{figure*}
\centering
\includegraphics[height=13cm,width=13.0cm]{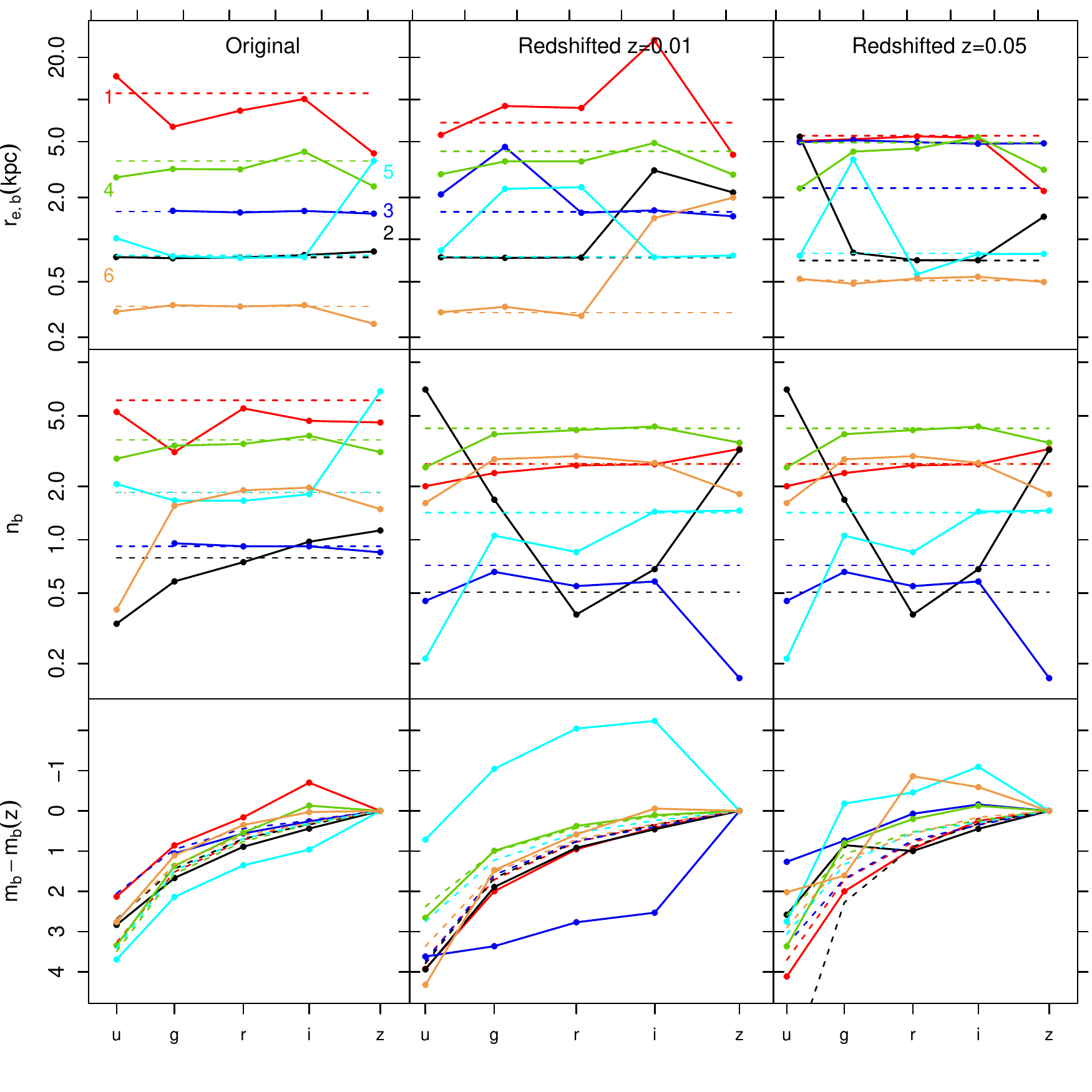}
\caption{Example results for six galaxies: 1-NGC2775 (red),  2-NGC4041 (black), 3-NGC4116 (blue),  4-NGC4365 (green),  5-NGC4638 (cyan) and 6-UGC08237 (orange).
We show the recovered effective radius of the bulge (top panels), bulge \sersic index (middle panels) and bulge SED (bottom panels) for both single-band (solid lines) and multi-band (dashed lines) fitting methods. 
The left column shows results from fitting our original SDSS images, while the other columns show results for images artificially-redshifted to $z=0.01$ and $0.05$. 
In our multi-band setup, the effective radius and \sersic index values are not allowed to vary with wavelength, while magnitudes have full freedom. Note that the NGC4116 $u$-band point is missing because the code crashed while fitting this single-band image. 
 We change the line style of the multi-band $n_{\rm b}$ orange lines to distinguish them from the overlapping lines.
The multi-band values are more consistent with increasing redshift, suggesting a similar improvement in stability as that found for single-\sersic fits in V13. 
}
\label{fig:sixgal}
\end{figure*}

In this paper we use the same set of 4026 galaxy images as in V13. These images comprise a sample of 163 nearby galaxies with imaging from SDSS in the $u$, $g$, $r$, $i$ and $z$ passbands. Our galaxies typically extend over more than one SDSS frame, so to create them we employ \montage \citep{tex:JK10}, which performs the transformations, rebinning and background adjustment necessary to combine the individual frames into a single mosaic.

In addition to the original images of our galaxies, we use 3863 further images in which the galaxies have been artificially redshifted.  We use \ferengi \citep{tex:BJ08} to create a set of $ugriz$ images mimicking the appearance that each of the 163 nearby galaxies would have if they were observed by SDSS at a range of redshifts.
The artificial redshifting algorithm applies cosmological changes in angular size, surface brightness and, optionally, shifting of the restframe passband (k-correction), to simulate the observation of a given galaxy at a greater distance. We produce images for redshifts $0.01$--$0.25$, in steps of $0.01$.  Note that not all our galaxies have images for every one of these redshifts, either because they originally have a redshift higher than $0.01$, or because the galaxy effectively becomes a point source.
As in V13, to avoid confusion with genuine redshift biases, we disable the k-correction feature of \ferengi in this work.

Full details of the redshifting process and the data preparation for both original and redshifted images have been given in V13.  In the present paper we use the same masks, PSFs and sky estimates. 

The morphological breakdown of our galaxy sample is given in the top row of Table~\ref{table:sample}.  All classes, except elliptical, also include barred types.  
Note that, while we ensure a broad range of morphologies are included, the distribution of Hubble types in this sample is not representative of the local Universe. As described in V13, our sample is comprised of galaxies which have had their structure carefully measured by previous studies.
We can then compare these with our semi-automated, multi- and single-wavelength results.

\begin{figure*}
\centering
\includegraphics[height=10.5cm,width=11.5cm]{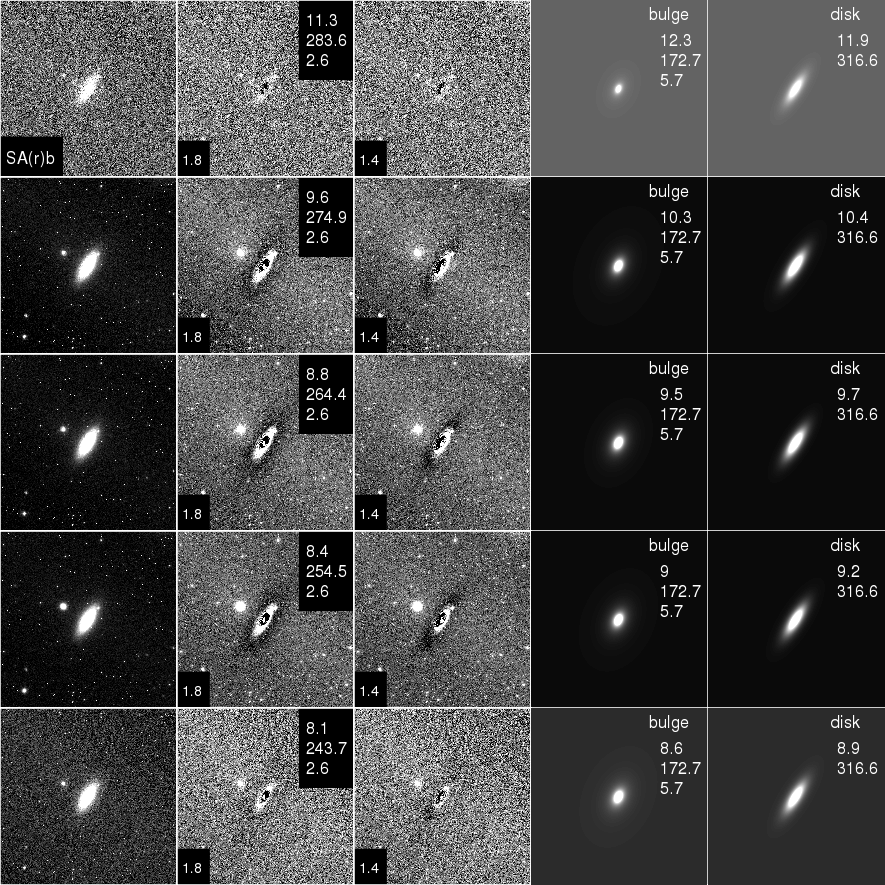}
\caption{Images of the galaxy NGC2841 in $u,g,r,i,z$ bands. The first column on the left shows the original images, the second column shows the residuals from the \fitc single-\sersic fit and the third column the residuals from the \fitc bulge plus disc fit. 
The fourth and fifth columns display the bulge model (\sersic function) and disc model (exponential function), respectively. In the second column the top-right legend gives the apparent magnitude, effective radius (in pixels) and \sersic index of the single-\sersic fit. The bottom-left legend in both the second and third columns gives the minimised $\chi^2$ of each fit as given by \galfitm. The legends in the fourth and fifth columns show the bulge and disc magnitude, effective radius (in pixels), and bulge \sersic index.}
\label{fig:modimag}
\end{figure*}

\begin{figure*}
\centering
\includegraphics[height=10.5cm,width=8.5cm]{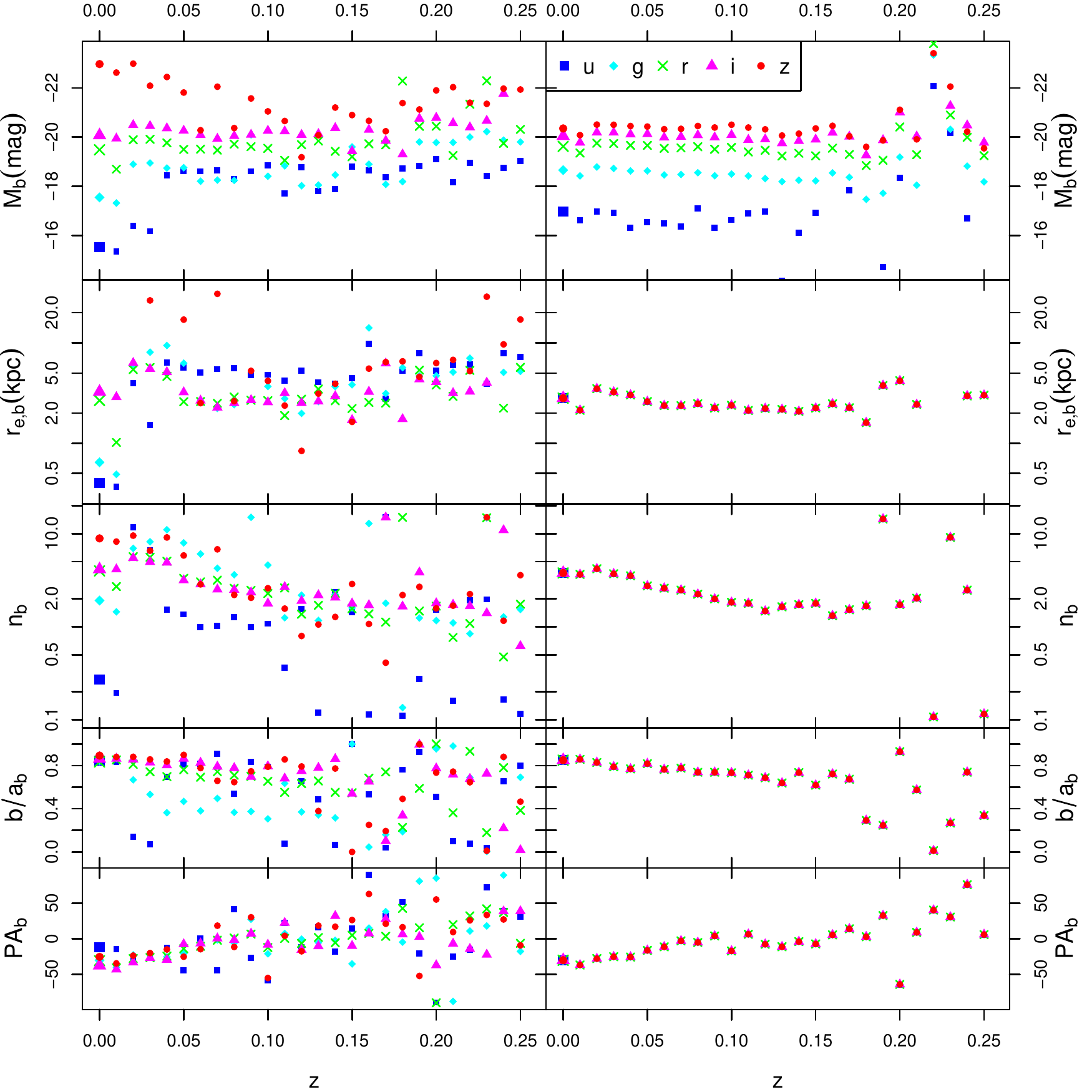}
\includegraphics[height=10.5cm,width=8.5cm]{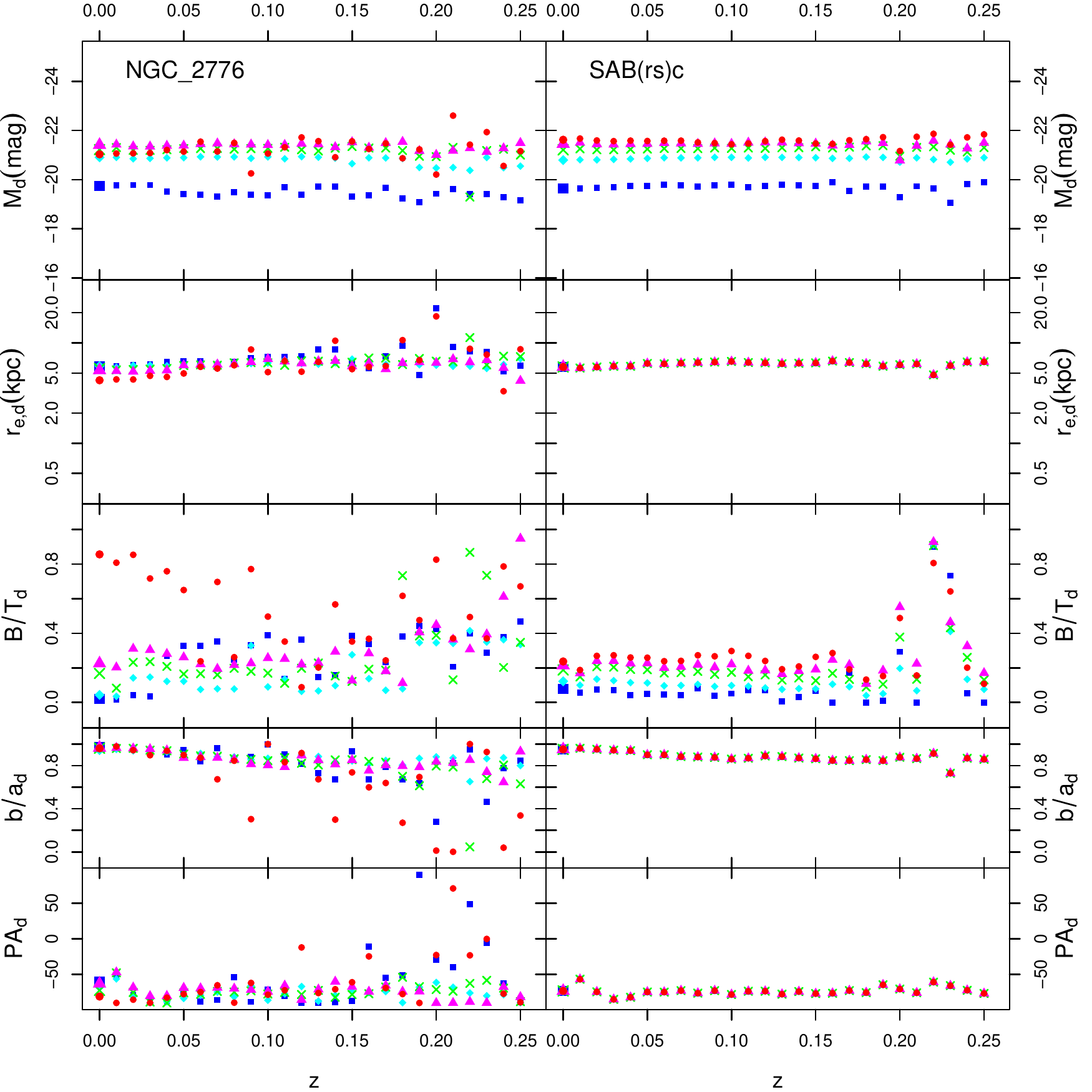}
\caption{A series of plots for galaxy NGC2776, presenting the variation of recovered parameters for the bulge (left panels) and the disc (right panels) as a function of redshift. Within each set of panels, the left column shows the single-band (\fitb) results, the right column shows the multi-band (\fitc) results. The points at redshift zero in each panel give the values for the original galaxy image, while the rest of the points represent the artificially-redshifted images. A different symbol is used for each band, as indicated in the legend. Note that in the panels showing disc properties, we plot the bulge-to-total flux ratio instead of the \sersic index, which is fixed to one.  Also note that the magnitude scales of the bulge and disc panels are different. 
}
\label{fig:gal1}
\end{figure*}

\begin{figure*}
\centering
\includegraphics[height=10.5cm,width=8.5cm]{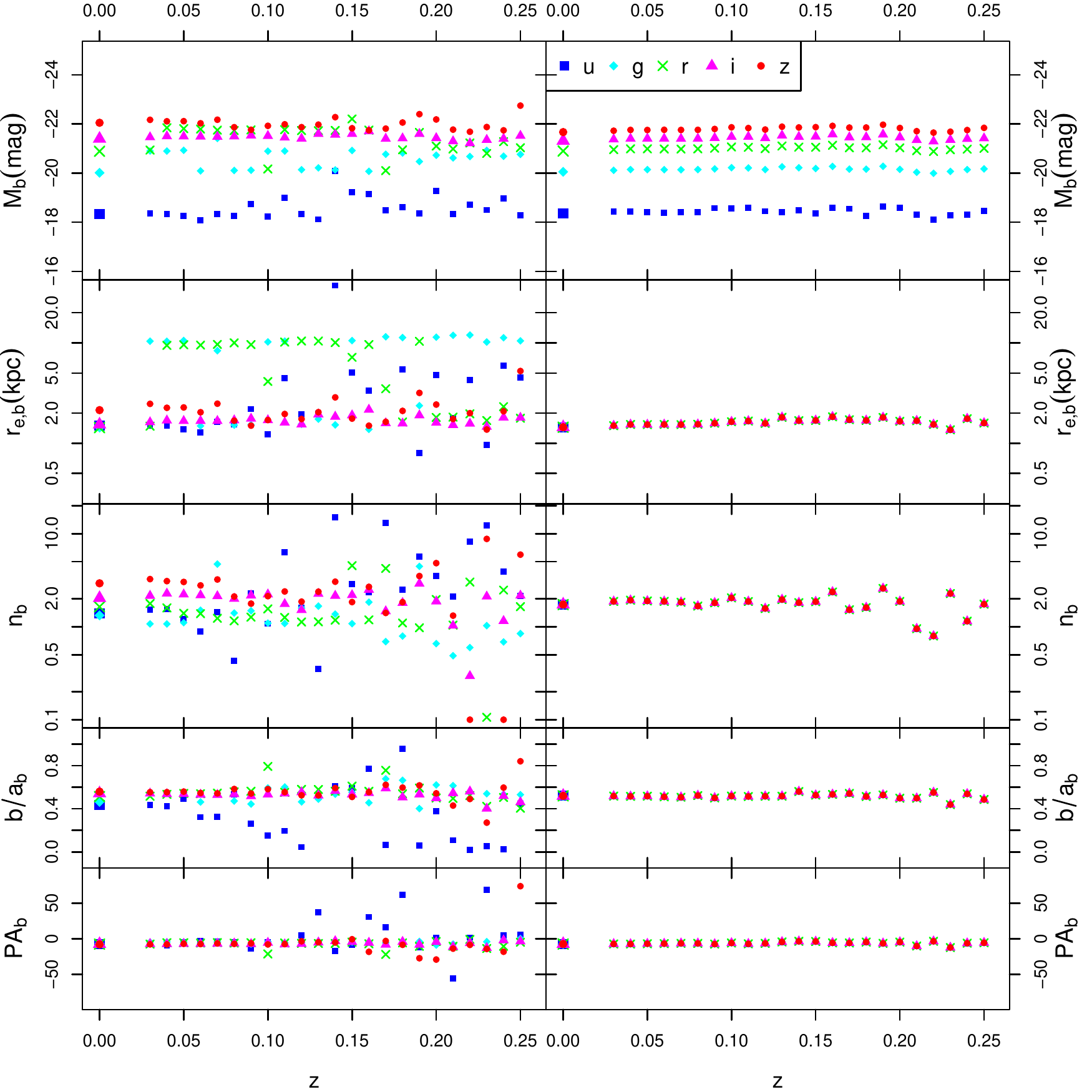}
\includegraphics[height=10.5cm,width=8.5cm]{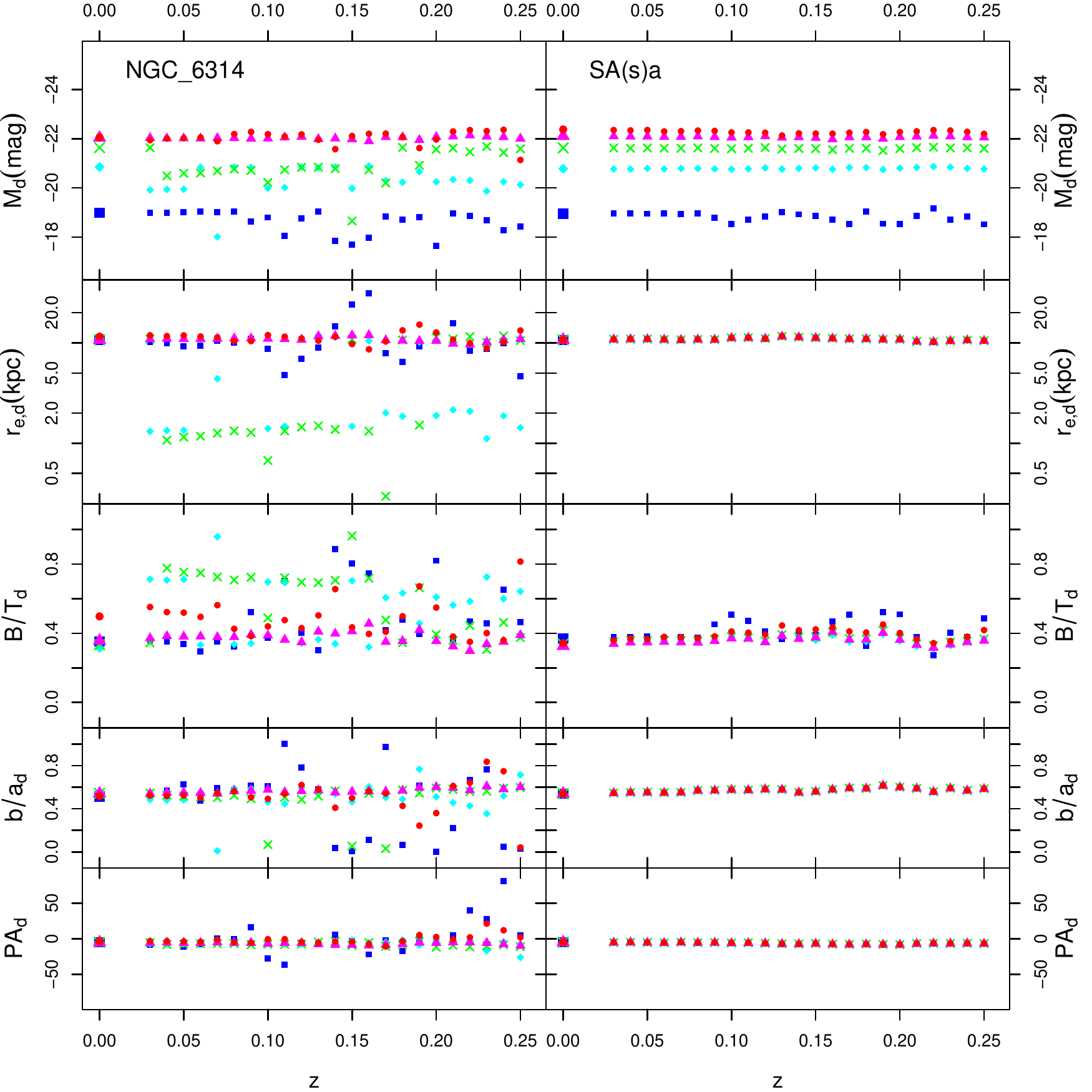}
\caption{A series of plots for galaxy NGC6314, presenting the variation of recovered parameters as a function of redshift in an identical manner to Fig.~\ref{fig:gal1}.}
\label{fig:gal3}
\end{figure*}

\subsection{Structural parameters}

\subsubsection{Fitting galaxies with \galfitm}

We use a modified version of \galfitthree to fit two-dimensional analytic models to our galaxy images.  A detailed description of \galfit is given by \citet{tex:PH02,tex:PH10}.
We have adapted \galfit (version 3.0.2) for the requirements of this project, as briefly described below.  To differentiate it from the standard release, we refer to our modified version as \galfitm\footnote{\galfitm is publicly available at \url{http://www.nottingham.ac.uk/astronomy/megamorph/}.}.  All the work in this paper uses version \galfitm-0.1.3.1.

Standard \galfitthree accepts only a single input image with which to constrain the model fit. To utilise multi-band data it was therefore necessary to make a number of significant modifications.  However, we have endeavoured to retain the original code unchanged, wherever possible.  \galfitm is therefore backward compatible and produces almost identical results to \galfitthree when used with single-band data (see section~4.1 in V13).

Our modified code can accept an arbitrary number of (pixel-registered) images of the same region of sky at different wavelengths. To these images, \galfitm fits a single, wavelength-dependent, model.  As for \galfitthree, this model may comprise one or more components, each with a number of parameters.  For example, for a single \sersic function, the parameters are: centre position ($x_{\rm c}$, $y_{\rm c}$), magnitude ($m$), effective radius ($\re$), \sersic index ($n$), axial ratio ($b/a$) and position angle ($\PA$).  To enable these model components to vary with wavelength, each of their standard parameters are replaced by functions of wavelength. For convenience, these are chosen to be Chebyshev polynomials (see \citet{tex:HB13} and Bamford et al. (in prep.) for more details).  Instead of directly fitting the standard parameters, \galfitm optimises the coefficients of these polynomials to best match all the multi-band data.

For each standard parameter, the user may select the order of the polynomial that describes its wavelength dependence, and thereby the freedom that parameter has to vary.  Some parameters may be entirely fixed; for example in this paper, we set $n = 1$ for the second \sersic function, in order to model an exponential disc component.  Other parameters may be allowed to vary as a constant with wavelength; e.g., one might allow the central $x$ and $y$ coordinates to vary during the fit, but require that they are the same in every band. Further parameters may be permitted to vary with wavelength as linear, quadratic, or higher-order functions; e.g., one could choose to allow the axis ratio $b/a$ to vary linearly with wavelength, in order to account for changes in ellipticity at different wavelengths.  Ultimately, specifying a polynomial with as many coefficients as there are input bands allows that parameter to vary freely with wavelength.
The user therefore has great flexibility to achieve a compromise between the freedom of the model, physical expectations, and the number of free parameters that must be constrained by the data.

A key element of our technique is that parameter values at the wavelengths of low signal-to-noise images can be partially interpolated or extrapolated from the higher signal-to-noise data.  
However, any significant signal present in those images should have an appropriate influence on the fit.  Systematic biases will be thus be reduced in comparison to an extrapolation based only on the high signal-to-noise bands.
The risk of such systematics may be further reduced by giving the model more freedom to vary with wavelength (e.g. linear or quadratic variation of $r_{\rm e,b}$, $r_{\rm e,d}$ and $n_{\rm b}$). The cost is an increase in the statistical uncertainties of the parameters in the low signal-to-noise bands.

In this work we have held most structural parameters fixed versus wavelength.  This corresponds to assuming a simplified picture of galaxy structure, in which galaxies comprise only bulge and exponential disc components, each without colour gradients and not departing from an elliptical projected \sersic profile.
Real galaxies may not obey these assumptions.
Therefore, while our structural constraints increase the stability of the fits, there is also a risk of introducing systematic biases in cases where the true wavelength dependence of the profile does not correspond to that assumed.  
Lower signal-to-noise images would be most susceptible to such systematics, as their parameters will be influenced by any higher signal-to-noise data.  

Our approach makes the assumed variation of galaxy structure with wavelength explicit, and allows one to relax these assumptions in a selective and gradual manner.
This flexibility allows the user to balance systematic and statistical uncertainties, using independent observational results, physical insight and knowledge of their dataset.

We plan to explore the variation of structural parameters with wavelength in detail in a future paper.
However, based on preliminary results we find that the vast majority of results present in this paper do not change by allowing small wavelength variation of the structural parameters.

\subsubsection{Model choices}

Most galaxies are considered to be primarily two-component systems, comprising a disc, with an exponential ($n=1$) profile, and a bulge, typically well represented by a \sersic function with $n\sim 0.5$--$4$ \citep{tex:G01}.
On the other hand, elliptical galaxies are generally considered to be single-component systems, describable by a single \sersic profile.

In V13 we performed single-\sersic fits to all of our images.  In this paper we supplement these with bulge-disc decompositions, performed using \galfitm to fit two superimposed elliptical \sersic models. 
For the first component we fit all standard parameters (effective radius $r_{\rm e}$, apparent magnitude $m$, \sersic index $n$, axis ratio $b/a$, and position angle $\PA$),
while for the second, we fix the \sersic index to be equal to one. 

In order to carry out a blind test of our analysis, and allow us to draw conclusions relevant for large surveys, we assume that we do not know the Hubble classification for our galaxies.  We therefore fit them all (even known ellipticals) with two functions.
In Section~\ref{sec:correlations}, we will explore what can be learned from this, including the potential for separating single-component systems, i.e. ellipticals and pure disc galaxies, from two-component systems, using structural parameters diagnostics.

We run two sets of fits, each of which is performed on the $u$, $g$, $r$, $i$, $z$ band images for all of our original and redshifted galaxies. For the first (single-band fitting; or \fitb, to reuse the nomenclature of V13) we fit each band individually. For the second (multi-band fitting; \fitc) we fit each galaxy using all five bands simultaneously.

We allow the magnitudes (for both bulge and disc) to vary completely freely between bands. For the multi-band method this amounts to setting the wavelength dependence of magnitude to be described by a quartic polynomial, with as many free coefficients as the number of bands.
We allow full freedom as we wish to avoid any potential biases on the recovered magnitudes, and hence colours, which may result from assuming a lower-order polynomial dependence. 

For the effective radius and \sersic index, we choose to not permit any variation with wavelength.  This effectively ignores colour gradients within each component, but keeps the overall number of free model parameters down, hopefully improving the reliability of the decomposition process.  
Any measurements of the wavelength dependence of individual components will be noisy and are unlikely to provide significant evidence to contradict the reasonable default position of a constant value.  This is therefore what we assume.
Our decision is supported by previous results in the literature. 
For instance, both \citet{tex:MC03} and \citet{tex:MC11} find that the \sersic index of the bulge, as well as the effective radius of the bulge and the disc, show no significant variation (or a slightly linear relation in rare cases) across optical and NIR wavelengths. 

In Figure \ref{fig:sixgal} we show our results (the effective radius, \sersic index and spectral energy distribution), for the bulges of six example galaxies fit in our original and artificially-redshifted images. 
For most of our galaxies, the results of the single-band fits (solid lines) show substantial fluctuations with wavelength, which worsen with increasing redshift.  The results of our multi-band fits (dashed lines), with $n$ and $r_{\rm e}$ constant with wavelength, recover reasonable values that are close to the average of the higher signal-to-noise bands ($gri$) for single-band fitting. 
The multi-band results are more resilient for different redshifts (e.g. black, red and blue lines in different columns). 

We also assume that the shapes of our galaxy components do not change with wavelength, so we set the axis ratio, position angle and galaxy centre to be constant with wavelength.

In both runs (\fitb, \fitc) we use the same initial parameters for galaxy center $(x_{\rm c},  y_{\rm c})$, magnitude $(m)$, \sersic index $(n)$, effective radius ($r_{\rm e}$),  position angle ($\theta$), axis ratio $(b/a)$ and sky background value (although different values are used for each galaxy image, see V13 for more discussion of the sky estimate).  We experimented with various different schemes for choosing initial parameters values, before selecting the following approach.

The sample of 163 original images is fit first. The initial magnitudes are determined using the \fitc single-\sersic model results found in V13. The initial value for the bulge magnitude was set to $m_{\rm ss} + 0.75$, where $m_{\rm ss}$ is the single-\sersic magnitude.  The initial disc magnitude was set to $m_{\rm ss} + 0.65$ in order to start with a slightly fainter bulge than disc.
The initial effective radius of the bulge was chosen to be $0.5 r_{\rm e,ss}$, where $r_{\rm e,ss}$ is the effective radius from the single-\sersic fit.  Similarly, the initial effective radius for the disc was set to be equal to the $r_{\rm e,ss}$
 We therefore use the observation that bulges are typically smaller than their host discs.  We found that starting with an equal bulge and disc there are more chances the bulge to fit parts of the disc component.

The initial \sersic index of the bulge was chosen to be $n_{\rm ss}$, while the \sersic index of the disc was fixed to unity, for an exponential profile. The initial values of disc axis ratio, disc position angle, disc and the bulge center ($x,y$) were set equal to the equivalent single-\sersic value. 
In cases where a parameter was variable with wavelength in the single-\sersic fit, but constant in the current paper, we took the median of the five values. 
Finally, the initial value of axis ratio for the bulge was set equal to 0.8, and the initial value of position angle for the bulge was arbitrarily set to 10 degrees. 
We confirmed that the final results of the fits do not depend on the these values.

All the parameters are allowed to vary during the fitting process, with the exception of the disc \sersic index and the sky background, which were kept fixed. \galfit and \galfitm give the option to constrain the range of values for each parameter in order to avoid unphysical results. We make use of this option by applying the following constraints. We require both magnitudes (bulge and disc) to vary within the range of $5$ to $35$ mag. Similarly, both the effective radius of the bulge and the disc were allowed to vary within the range of $0.04$ to $600$ arcsec. We constrain the freedom  of the \sersic index by allowing it to vary within the range of $0.1$ to $15$. However, we exclude any bulge with $n_{\rm b}<0.3$ from the final sample. Finally in the case of the center ($x$, $y$) we applied two constraints, one to fix the bulge and disc to have the same center, and a second to restrict their variation, with respect to the single-\sersic fit, to be no more than $\sqrt{s/8}$, where $s$ is the size of the image.

For each artificially-redshifted image, we repeat the same procedure as above to estimate the initial parameter values, but use the \fitc single-\sersic result obtained for the same redshifted image. 
In cases where a single-\sersic magnitude was unphysically faint, we calculated the initial parameter values by cosmologically adjusting the values obtained for the lowest-redshift artificial image. 
We apply the same constraints as for with the original galaxies.

In addition to \fitb and \fitc, we perform another set of fits to the artificially-redshifted images, which we refer to as `aperture fits'. For these we take the structural parameters from the \fitb $r$-band results and keep these fixed while performing single-band fits to the $u$, $g$, $i$ and $z$ band images.
Only the magnitudes are allowed to vary freely during the fit. In this way we apply an identical model in all the bands and ensure we only measure the variations in the flux for a fixed `aperture'.  This approach is commonly applied to ensure meaningful colours.

\begin{table}
\caption{The number of galaxies with reliable bulge and disc components, divided by morphology and waveband.}
  \smallskip
  \centering
  \begin{tabular}{ccccccccc|c}
  \hline
  \noalign{\smallskip}
& Band & E  & S0  & Sa   & Sb  &    Sc  &      Sd       &     Sm/Irr  & Total   \\ 
\noalign{\smallskip}
\hline
\hline
\noalign{\smallskip}       

Total &  & 23 & 18  &8  &29 & 50 & 24 & 11 &163 \\
 \noalign{\smallskip}
\hline
 \noalign{\smallskip}

                   &  u     &     23     &     17      &   7   &       18      &     30       &   11      &     6    &  112 \\
Reliable    &  g     &     23     &     17      &   7   &       20      &     32       &   11      &    8    & 118 \\
bulge         &  r      &     23     &     17      &   7   &       21      &     32       &   12      &   8   &  120 \\ 
                    &  i      &     23     &     17      &   7   &       21      &     32       &    13     &    8  &  121 \\  
                    &  z     &     23     &     17      &   7   &       21      &     32       &    14     &    8  &  122 \\ 
 \noalign{\smallskip}
\hline
 \noalign{\smallskip}
                &  u     &   19       &  17     &  7  &   27  &   49    &  23  &   11  &  153  \\
Reliable &  g     &    19      &  17     &  7  &   27   &   49     & 23   &  11   &  153  \\
disc         &  r      &   19       &   17    &  7  &  27   &   49    &  23  &  11   &   153 \\ 
                &  i      &    19      &   17   &  7  &    27   &  49      &  23 &  11   &   153  \\  
                &  z     &     19     &   17   &  6  &    27   &   49     & 23  &  11   &   152  \\    
 
  \noalign{\smallskip}
\hline
 \noalign{\smallskip}

                   &  u     &    19    &   16      &   7   &   18     &  30      &  11     &    6     & 107 \\
Reliable    &  g     &    19    &    16    &   7   &   20     &  32      &  11     &    8     & 113 \\
bulge         &  r      &    19    &    16     &   7   &   21     &  32      & 12     &    8   & 115 \\ 
and disc    &  i      &    19     &    16     &   7   &   21     &  32      &  13    &    8   & 116 \\  
                   &  z     &    19     &    16     &   6   &   21     &  32      &  14    &    8   & 116 \\    
\noalign{\smallskip}
\hline
\end{tabular}
\label{table:sample} 
\end{table}

\subsection{Inspection of individual galaxies}
\label{sec:inspection}

Figure~\ref{fig:modimag} shows the original $ugriz$ images, residuals from the  single-\sersic and bulge-disc \fitc fits, and the separate \fitc  bulge and disc model components, for an example spiral galaxy 
The figure also includes various useful numbers for each fit.  Using similar figures we have visually examined the fitted models -- and their residuals -- for all of our 163 galaxies, to ensure that their shape and size correspond to the real galaxy.

In addition to checking the images, we also inspect all the recovered parameters for both the original and the artificially-redshifted images for each galaxy. 
In a similar manner to V13, Figs.~\ref{fig:gal1} and \ref{fig:gal3} present a summary of the bulge and disc results for two example galaxies.  
Equivalent plots are available for all the 163 galaxies.  The left panel shows the \fitb results and the right panel the \fitc results. 
At redshift zero we plot the results from the original images. The first row of panels shows the absolute magnitude ($M$), the second row shows the effective radius, the third row shows the \sersic index in the case of the bulge panel and the bulge-to-total flux ratio in the case of the disc panels. 
The last two rows show the axis ratio and the position angle. 

In these figures we determine the absolute magnitude and the effective radius assuming distances simply derived from the observed redshift and adopted cosmology. 
Therefore, the values shown for the original images in Figs.~\ref{fig:gal1} and \ref{fig:gal3} could differ slightly from later figures, for which we use more directly determined distances when they exist.

Figs.~\ref{fig:gal1} and \ref{fig:gal3} illustrate some of the behaviours seen for our fits as the galaxies are simulated as they would appear at higher redshifts.  In Fig. \ref{fig:gal1} we present the recovered structural parameters for the galaxy NGC2776. 
We can see that the single-band results fit different structures in each band, particularly for the bulge, and that the parameters of these structures vary substantially with small changes in the simulated redshift.
In contrast, the multi-band results are much more stable as a function of redshift, although some small trends are seen before even these results become noisy and unreliable at $z \ga 0.15$.  The systematic decline in \sersic index with simulated redshift appears to be a consequence of diminishing spatial resolution, and was also a fairly common feature of the single-\sersic component fits in V13.  For both methods \fitb and \fitc the disc parameters are much more stable than those of the bulge, presumably due to the disc's larger size and less steep inner profile.
Note that, even though the bulge and disc magnitudes are completely free to vary between bands, constraining the wavelength variation of $n$, $r_{\rm e}$, $\PA$ and $b/a$ through multi-band fitting leads to much more stable measurements of the magnitudes, and hence colours.

Figure \ref{fig:gal3} presents another set of  recovered structural parameters, this time for the galaxy NGC6314. The first thing to notice is that, for the single-band fit, the effective radius of the bulge is much larger than the disc for the $g$- and $r$-bands. 
This is an indication that, in these bands, the \sersic function is fitting the disc and the exponential function fitting the bulge, especially given the $n$ behaviour. 
The usual solution to this problem is to apply constraints on the fit, e.g. insist that the bulge be smaller than the disc.  However, using the same set of constraints for the whole sample may introduce biases in other galaxies.  Turning to multi-band solves this problem without requiring constraints, now the same structural components are fit in all the bands.
In addition, again we see a reduction in the variations caused by small changes in simulated redshift, and that the fit remains more reliable to higher redshifts.

So far we have shown, via specific examples, that that our multi-band approach can measure the fluxes and sizes of galaxy bulge and disc components more reliably than if each band is fit individually, at least when we allow no freedom for the $r_{\rm e,b}$, $r_{\rm e,d}$, $n_{\rm b}$, $b/a_{\rm b}$, $b/a_{\rm d}$, $\PA_{\rm b}$ and $\PA_{\rm d}$ parameters to vary with wavelength.  
Substantial variations in the recovered parameters with relatively small changes data quality (redshift) are dramatically reduced.
The improved stability is particularly noticeable at low signal-to-noise (S/N). 
As a result, it increases the distance out to which meaningful bulge-disc information can be recovered for a galaxy of a given luminosity.  
In Section~\ref{sec:stat_trends} we demonstrate these improvements in a more general manner, by considering the average trends of various parameters versus redshift, for our whole galaxy sample.

\begin{figure}
\centering
\includegraphics[width=0.45\textwidth]{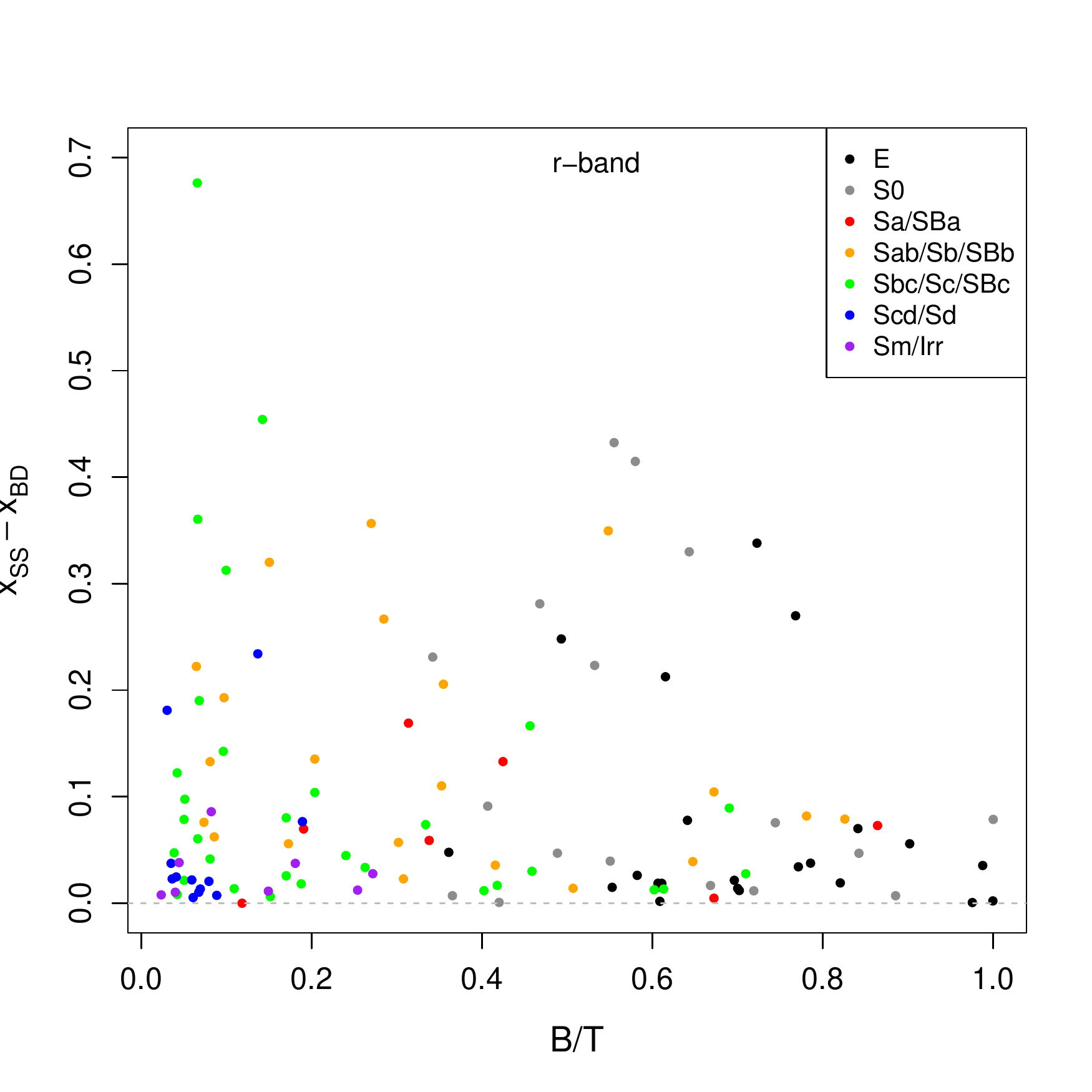}
\caption{The difference in $\chi^2$ between the single-\sersic and the bulge-disc fits, as function of bulge-to-total flux ratio. Only galaxies with a significant bulge are shown in this figure.}
\label{fig:chisq}
\end{figure}

\begin{figure}
\centering
\includegraphics[width=0.5\textwidth]{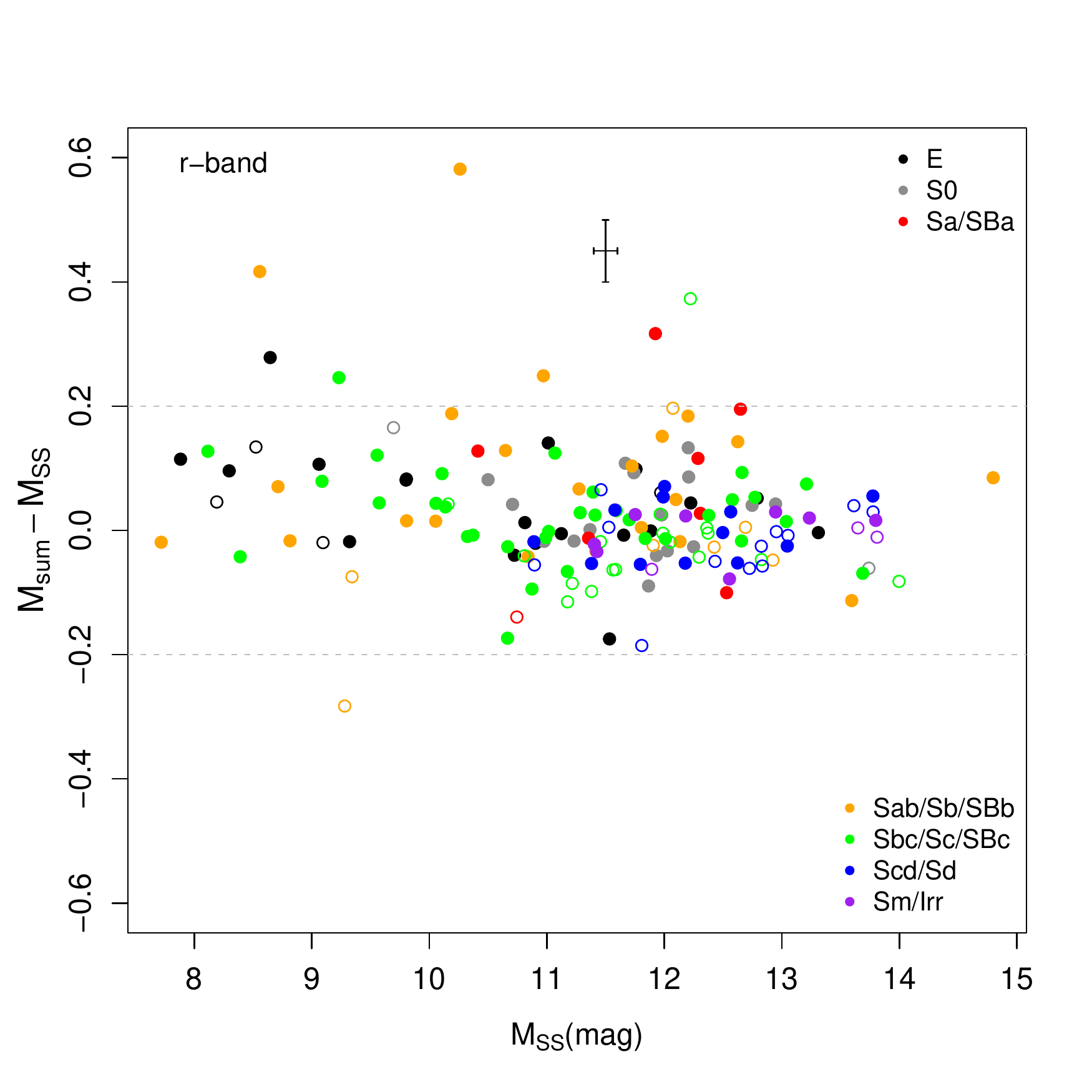}
\caption{Comparison of the single-\sersic total magnitudes against  the sum of the component magnitudes from our bulge -disc fits. Filled circles denote galaxies with both a reliable bulge and disk. Representative error bars for our measurement are displayed in the top part of each panel. See text for further discussion on the uncertainty measurements.}
\label{fig:ssbd_tmag}
\end{figure}

\subsection{Obtaining reliable structural measurements}
\label{sec:sample}

Before studying the distribution of the galaxy component parameters, we must select a sample with reliable bulge-disc measurements. 
In this section we describe the various controls we apply to determine if our fits are physical meaningful. 
For those galaxies where the fitted model is a poorly match to the original image, given our physical expectations, we repeat the fit again with different initial parameter values or additional model components.  
We aim for our procedure to be applicable in an automated manner, that could be used for large surveys. 
However, as we have a small sample of galaxies, we still use the tools described in Section~\ref{sec:inspection} to inform our choices, to check if the automated selection agrees with visual inspection, and to aid the interpretation of our results. 

We start by identifying unphysical models where re-attempting the fit may produce a better outcome. Examining the results of the multi-band fits to the original images, we find that 50 galaxies (out of 163) have bulges with $r_{\rm e,b}/r_{\rm e,d}>0.9$.  
After further investigation we separate these galaxies into two cases.
The first group consists of 17 galaxies for which the $n=1$ component is fitting the inner structure of the galaxy and the free-$n$ component is fitting the disc. 
For these the measured \sersic index varies between $0.5$ and $1.4$. All these galaxies are late type spirals (Sc, Sd, Sm) and 13 of them have a bar.
For these cases we believe that the initial parameter values were far from optimal. 
We choose to fit these galaxies again, using a different set of initial values, with a brighter flux and larger size for the intended disc and lower flux and smaller size for the bulge. The new fit corrects the problem for the vast majority of the 17 galaxies.

The second group consists of 33 galaxies where the bulge fits an inner structure, but also dominates the outer region of the galaxy. In this group we find 10 early-type galaxies (E/S0), where 9 have bulge \sersic values $2.8$--$5.7$ and one, NGC4458, has $n_{\rm b}=11.3$. 
The remaining 23 galaxies are spiral galaxies (Sa, Sb, Sc) with bulge \sersic values between 4 and 11, usually accompanied by a high bulge-to-total ratio. 
These  $n_{\rm b}$ and $B/T$ values are unusually high for late type spirals. For this second group of galaxies, we initially re-fit the galaxies in the same way as for the first group. 
This approach corrects the fits for almost one-third of the cases. 
For the remaining galaxies we attempt another fit with an addition of a third component (in the form of a central point source), together with the second set of initial parameters. 
We choose to add a PSF function to account for any extra flux in the centre of these galaxies that could be responsible for the high values of \sersic index and effective radius.

These new fits return smaller bulges with lower \sersic indices for another one-third of the cases, and they reduce the \sersic index without reducing the effective radius for a further four galaxies.  
For the remainder we do not adopt the fit results with additional-PSF component, either because the value of the bulge effective radius or \sersic index was larger than before, or the PSF magnitude was negligible ($< 30$ mag). One case where the addition of a PSF component failed to improve the fit is the elliptical galaxy NGC4458 (see the Appendix~\ref{sec:notes} for more details on this galaxy).

\begin{figure*}
\centering
\includegraphics[width=0.85\textwidth]{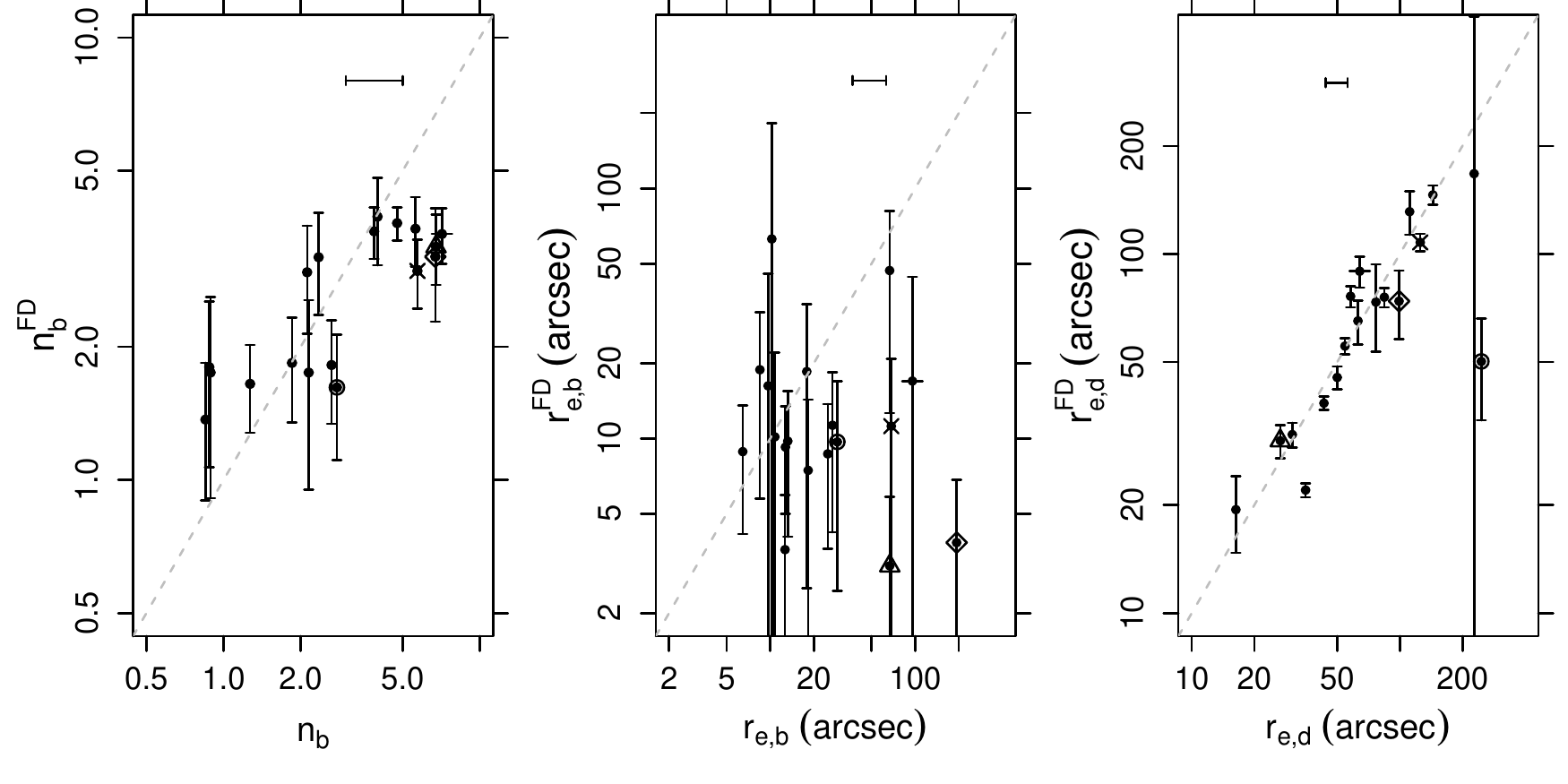}
\caption{A comparison between the structural parameters derived in this work and those obtained by \citet{tex:FD08} for 18 galaxies we have in common.
The first panel displays the difference in bulge \sersic index.  The second and third panels show the effective radius of the bulge and disc, respectively.
Symbols: x - NGC2841, rhomb - NGC3521, triangle - NGC3642, cross - NGC4698, circle - NGC4736.
Representative error bars for our measurement are displayed in the top part of each panel.  See text for further discussion on the uncertainty measurements.
}
\label{fig:FD08}
\end{figure*}

After refitting both groups we then select our final samples of trustworthy bulges and discs.
We reject bulges in the multi-band, original-image results with fit parameters on any of the constraints, insignificant bulge components (at least 3 mag fainter than the disc or below the SDSS point source detection limit), and bulges with effective radius smaller than 5 pixels.
We perform these checks in each individual band.
The final numbers of galaxies with acceptable bulge measurements are shown in Table~\ref{table:sample}. A large fraction of late-type spiral galaxies do not possess a reliable bulge measurement. 
For all these cases we trust the disc parameters but exclude the bulge properties from further analysis.

In the case of the disc component, we find a few occasions where the effective radius of the disc has taken unreasonably large values or its total brightness is more than three magnitudes fainter than the bulge. 
For these cases the bulge component returns very similar results to the single-\sersic fit (Paper II), the disc only accounts for a minor details in the residual. 
Four of these galaxies (NGC4360, NGC4378, NGC4486, NGC4621) are elliptical galaxies and consequently may indeed lack a disc. 
In addition, we fail to fit a disc component for NGC4459, even though it is classified as S0. For these objects we trust the bulge parameters but exclude the disc properties from further analysis. 

Two problematic cases that were discovered through the above checks are the galaxies NGC4378 and NGC4450. For these two galaxies both the bulge and disc components have been removed. While both these galaxies are two component galaxies, some image distortions hinder the bulge-disc decomposition process.

Around 110 galaxies have both bulge and disc components that we deem as trustworthy, depending on the band.
We should stress that this sample includes 19 elliptical galaxies that were fitted with two functions. 
For these galaxies, we do not know if these two components correspond to truly distinct structures with different kinematics, but we choose to keep them in the further analysis.  So far we have not found any indication that their fits are inappropriate, other than their visual classification\footnote{The visual classifications have been taken from NED}, which would not be available for a large automated sample.
The number of galaxies that have both trustworthy bulge and disc measurements are broken down by morphology and band in Table~\ref{table:sample}. 

We attempt to identify elliptical galaxies by comparing the goodness-of-fit of our one- and two-component fits. 
In Fig. \ref{fig:chisq} we plot the difference in reduced-$\chi^2$ between the single-\sersic and bulge-disc fits as a function of bulge-to-total flux ratio. 
The improvement in reduced-$\chi^2$ for the elliptical galaxies is in the same range as the other Hubble categories. 
We find that 18\% of our elliptical galaxies and 33\% of our lenticular have a dramatic reduction ($\chi_{SS}^2-\chi_{BD}^2>0.1$) in their reduced-$\chi^2$  by adding an extra exponential function. 
Only 13\% of our elliptical galaxies and 17\% of the S0 show a negligible change in reduced-$\chi^2$ ($\chi_{SS}^2-\chi_{BD}^2<0.01$). 
In cases where a PSF function has been included we use the $\chi_{BD}^2$ for that fit.
The addition of the PSF function in all the cases improved the $\chi_{BD}^2$ by less than $0.01$.

Chi-squared should always decrease when a model is given more freedom. The Bayesian Information Criterion (BIC) is based on $\chi^2$, but penalises additional parameters, in an attempt to provide a guide to whether the additional freedom is warranted by the data.
However, applying the BIC to our data finds only nine cases where the single-\sersic fit is deemed better than the bulge-disc model. 
All these cases have already been identified as having an insignificant bulge or disc by our above selection criteria. We conclude that, by using the reduced-$\chi^2$ or BIC, we cannot select a clean sample of elliptical galaxies.

Quantifying the uncertainties of the bulge and disc structural measurements is a challenging task. 
In V13 we provided the following uncertainties for our single-\sersic fits: ($u$, $g$, $r$, $i$, $z$) for $m$ ($\pm0.13$, $\pm0.09$, $\pm0.1$, $\pm0.11$, $\pm0.12$), $r_{\rm e}$ ($\pm12$\%, $\pm11$\%, $\pm12$\%, $\pm14$\%, $\pm15$\% ) and $n$ ($\pm9$\%, $\pm11$\%, $\pm14$\%, $\pm15$\%, $\pm17$\%).
These were based on plausible systematic uncertainties in the sky estimation, which typically dominates the error budget.
As bulge-disc decomposition is a more complicated task, we expect the uncertainties on our bulge and disc measurements to be even larger. 
Overall, as we will also see from the further analysis, the bulge parameters are more dependent on the initial conditions, while the disk parameters are more robust. 
We attempt to determine indicative uncertainties on our fit parameters by refitting a randomly selected sample of 10 galaxies with different sky values. We alter the sky values by our estimated systematic sky uncertainties, as before.
We find that both $n_{\rm b}$ and  $r_{\rm e, b}$ can change by up to $\sim 25$\%.  The parameters of the disc are less strongly affected and are on the same level as the single-\sersic uncertainties.

\begin{figure}
\centering
\includegraphics[width=0.47\textwidth]{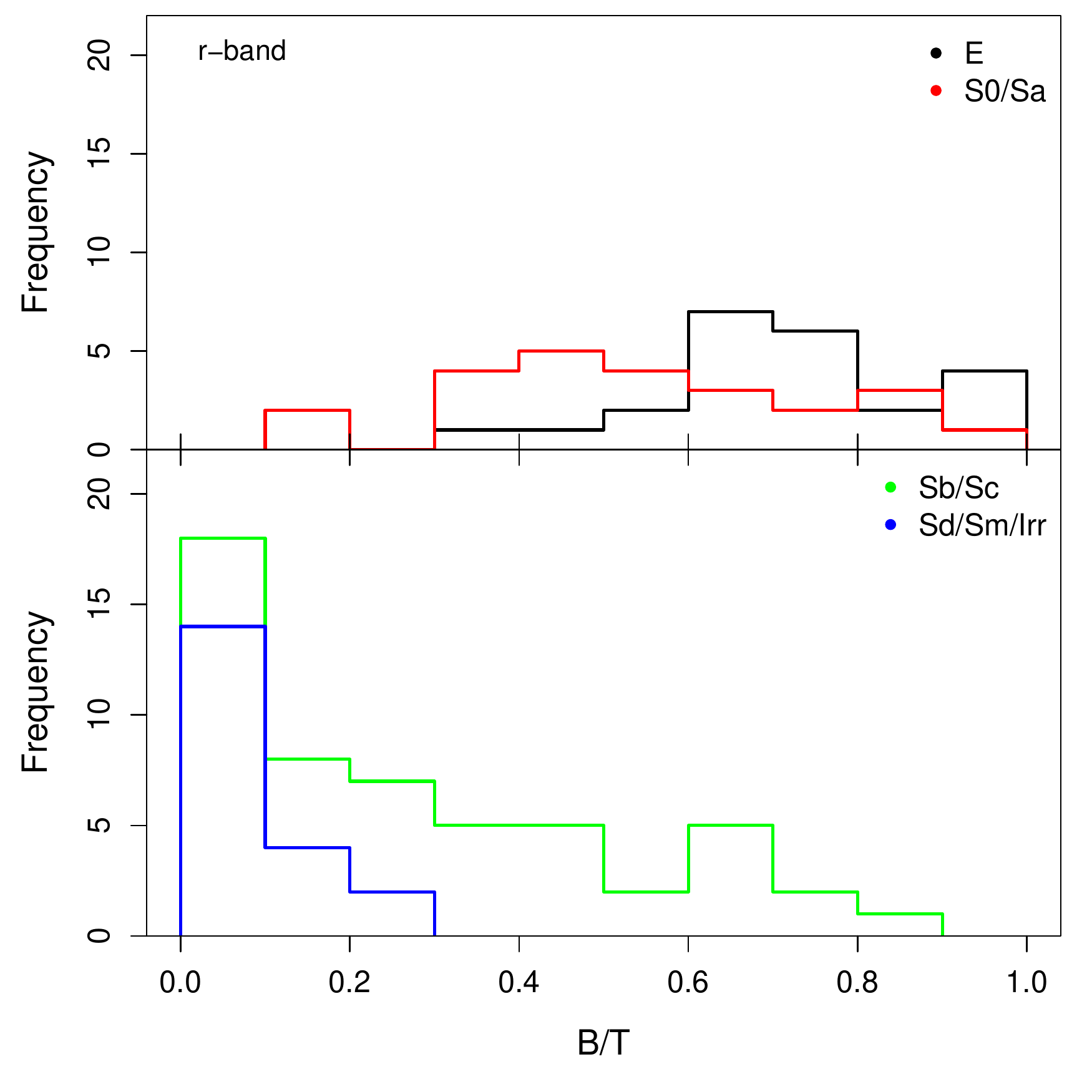}
\caption{The distribution of bulge-to-total flux ratio for different morphological bins, measured in the $r$-band using our multi-band (\fitc) method. Only galaxies with a significant bulge are shown in this figure.}
\label{fig:BThist}
\end{figure}

\begin{figure}
\centering
\includegraphics[width=0.47\textwidth]{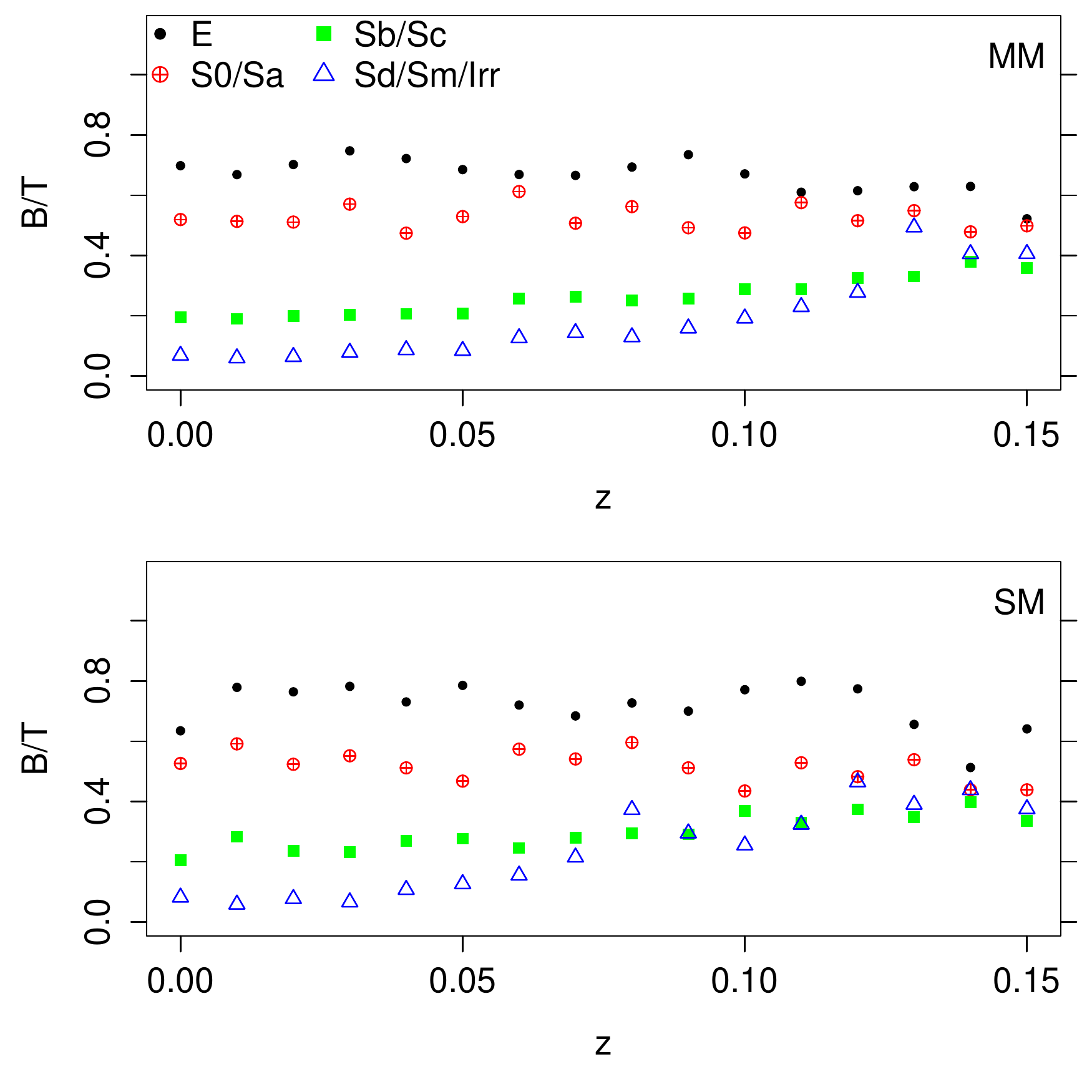}
\caption{The median value of the bulge-to-total distribution, as seen in Fig.~\ref{fig:BThist}, as a function of apparent redshift. At redshift zero the results from the original images are plotted, while for higher redshifts we show the results from the artificially-redshifted images. 
}
\label{fig:BTz}
\end{figure}

As a simple check of our fits, Fig.~\ref{fig:ssbd_tmag} shows the difference in the recovered $r$-band total magnitude between our single-\sersic and two-component models. We colour-code the galaxies based on their Hubble classification. 
For the vast majority of the galaxies the difference between the total magnitudes is smaller than 0.2 mag.
Bright galaxies $m<11$ show a small systematic trend, where the single-\sersic magnitude ($m_{\rm ss}$) is brighter than  the sum of the bulge and disc magnitude ($m_{\rm sum}$).
The outliers in this plot are NGC5850, NGC4725, IC0724, NGC4636, NGC4303 and NGC5806 with  $M_{\rm sum}-M_{\rm ss} = 0.58, 0.42, 0.32, 0.28,  0.25 $ and $ 0.25$ respectively.

We compare our derived bulge and disc parameters with those measured by previous studies, specifically those that analysed small samples of galaxies and carefully fitted each galaxy individually. 
Initially, we compare with  \citet{tex:FD08}, which presents structural parameters for 18 galaxies common to our sample, using $V$-band imaging from various sources. 
For comparison we convert our magnitudes to $V$-band central surface brightnesses.
For the remaining parameters, $r_{\rm e,b}$, $r_{\rm e,d}$ and $n_{\rm b}$ we compare our multi-band results, which do not vary with wavelength, directly with the V-band measurements from \citet{tex:FD08}. 
Figure~\ref{fig:FD08} shows these comparisons. 
The agreement is satisfactory for most of the structural parameters. Both sets of bulge measurements display large error bars. 
The discrepancy between the two different studies appears to be a result of the method used to model the galaxy (1D versus 2D) and of different masking theme. Fisher \& Drory decomposition is based on the major axis of each galaxy while ours utilises all the spatial information of the image. Various studies, e.g. \citet{tex:FD04,tex:PH10}, have found that the parameters derived with 1D fitting methods and 2D are not always in agreement. Even different approaches of 1D fitting e.g. major vs minor axis fitting can change the results. The second possible reason is that we fit the entire galaxy while in \citet{tex:FD08} they manually exclude the inner and outer part of the galaxy. This practice has the advantage that you can exclude for instance the inner part of the galaxy that may not follow the \sersic function but has the disadvantage that the fitting process dependents on personal choices (or personal experience) of what should be included in the fitting and what should be excluded.
The four galaxies that show significant offset are discussed further in Appendix~\ref{sec:notes}.

Additionally, we compare our results with \citet{tex:M04} for our 12 common galaxies, some of which are also in the \citet{tex:FD08} comparison. 
In order to perform a sensible comparison, we first recalibrated the absolute magnitudes found in \citet{tex:M04} to the distances used in this paper, and the AB zeropoint system. 
Finally we converted $UBVRI$ magnitudes to $ugriz$ using the transformations provided by \citet{tex:BR07}. 
Our results are similar to \citet{tex:M04}, with $\Delta M < 0.5$ mag, except for the u-band measurements of NGC2841, NGC3521, NGC4274 and NGC4303. Further information about some of these cases can be found in Appendix~\ref{sec:notes}.

\section{Structural properties versus morphological type and redshift}
\label{sec:stat_trends}

Our primary aim in this paper is to demonstrate that our multi-band decomposition method is able to determine physically meaningful bulge and disc parameters, both for nearby galaxies with high-quality imaging and more distant galaxies with noisier and less well-resolved images.  
In this section we therefore study the behaviour of various bulge and disc parameters for galaxies with different morphological types.  
For each parameter we first present the distributions as measured on our original SDSS images, using our multi-band (\fitc) approach. 
We do not show \fitb results for the original imaging because at such high resolution both methods return similar results (see V13 and below).  
We then investigate the stability of our measurements on the artificially-redshifted images, by examining how the median parameters for each morphological group vary with redshift, and comparing the multi-band (\fitc) and single-band (\fitb) methods.

\begin{figure}
\centering
\includegraphics[width=0.48\textwidth]{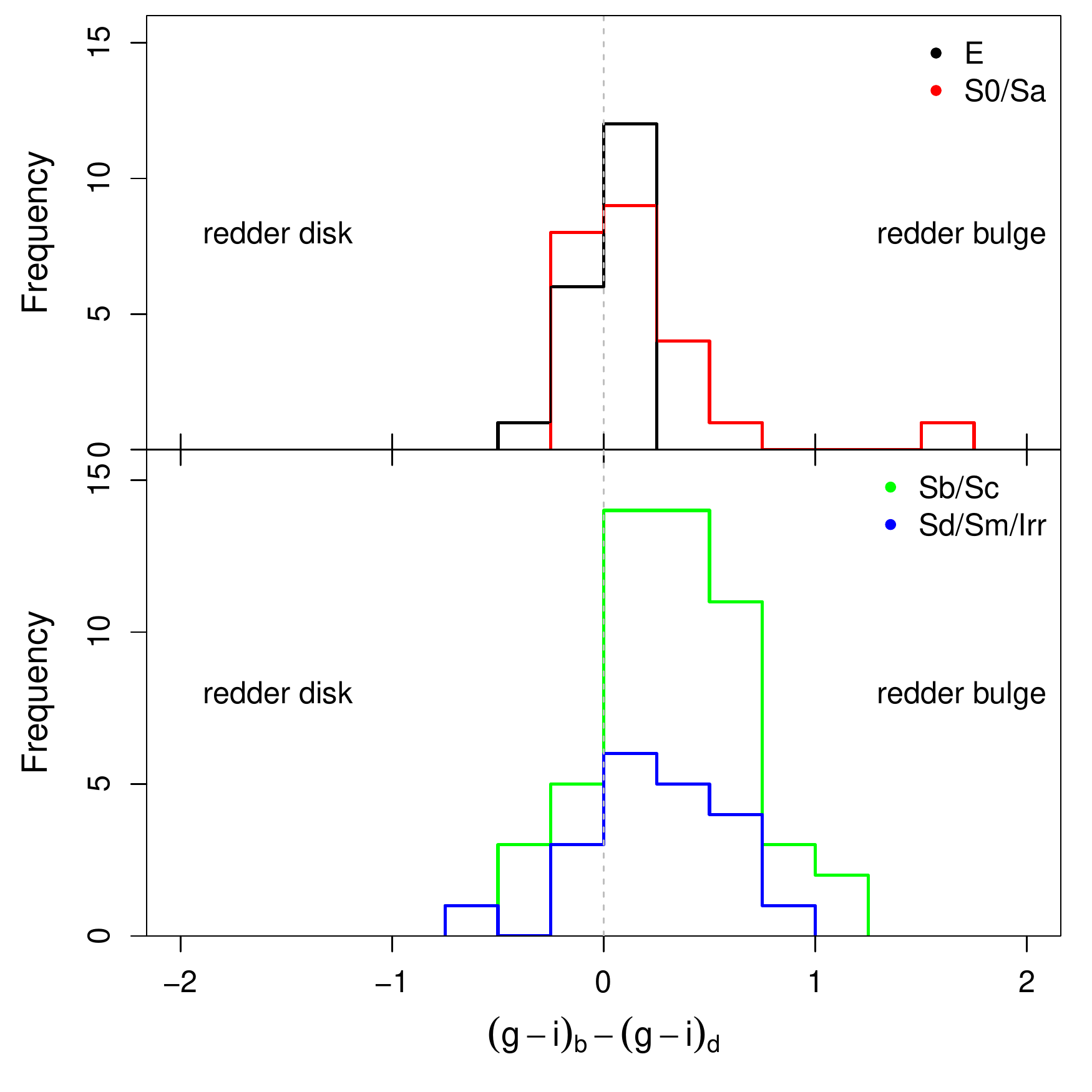}
\caption{The distribution of the bulge$-$disc colour difference for the galaxies measured in the original imaging. Only galaxies with both a significant bulge and disc are shown in this figure.}
\label{fig:coldiffm}
\end{figure}

\begin{figure}
\centering
\includegraphics[width=0.48\textwidth]{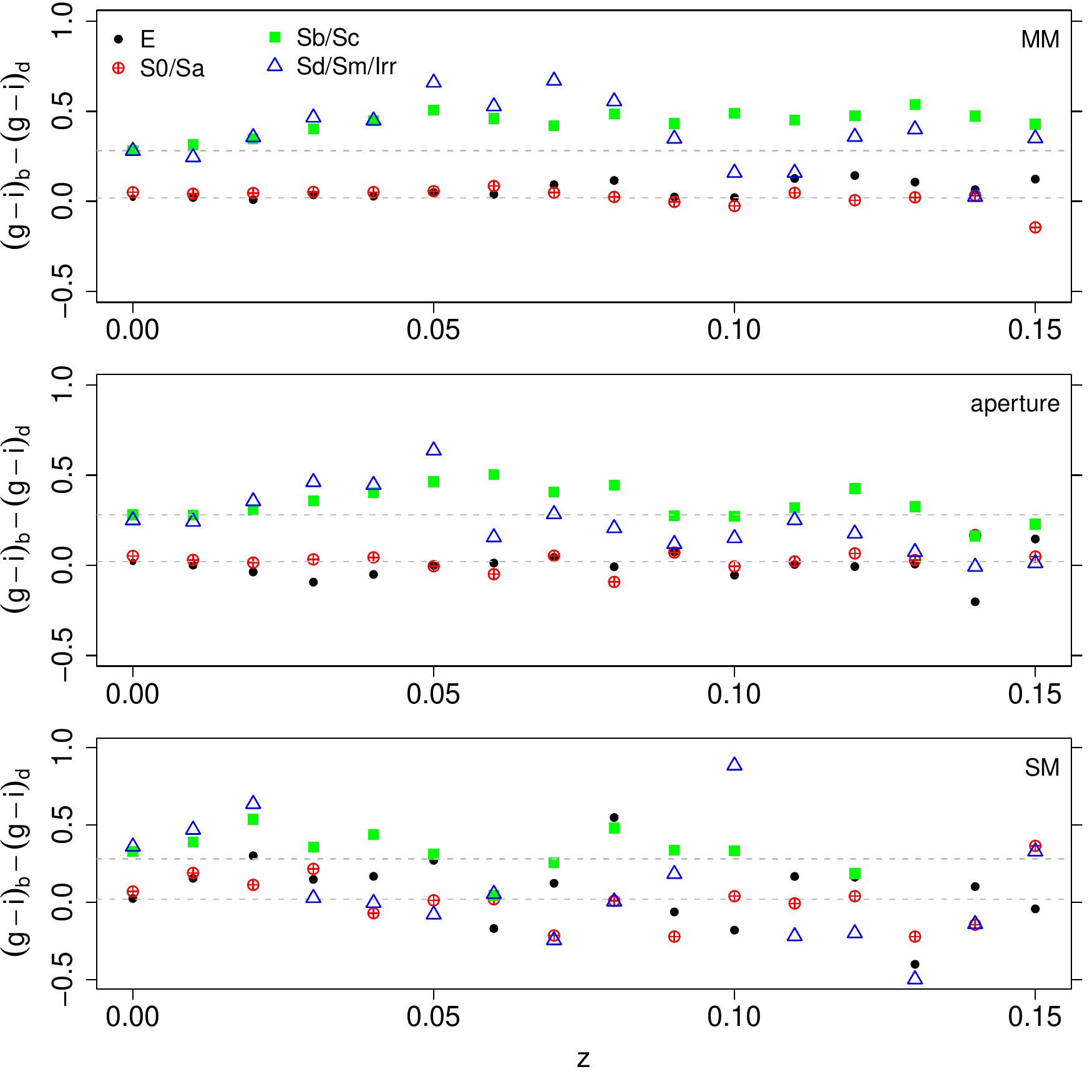}
\caption{The median value of the bulge$-$disc colour difference distribution, as seen in Fig.~\ref{fig:coldiffm}, as a function of redshift. 
At redshift zero the results from fitting the original images are plotted, while for higher redshifts we use the results from the artificially redshifted imaged.
The horizontal lines are the median values of the E and Sb--Sc groups, as measured in the top panel.
}
\label{fig:coldiff}
\end{figure}

In Figures \ref{fig:BThist}--\ref{fig:coldiffm} and \ref{fig:coldiff}--\ref{fig:r_ferengi} we divide our galaxies into four morphological groups: E, S0--Sa, Sb--Sc, and Sd--Irr.  Where we present the results of artificial redshifting, we only plot up to a redshift of $0.15$ (in contrast to $0.25$ in previous figures), as beyond this redshift neither approach produces useable structural measurements.

\begin{table*}
\caption{The average $r$-band structural parameters of our sample in different bins of Hubble-type. The errors quoted are the uncertainty of the median values. The additional row for Sb--Sc types shows the result of excluding the four galaxies with $r_{\rm e,b}$/$r_{\rm e,d}>1$.}
  \smallskip
  \centering
  \begin{tabular}{ccccccc}
  \hline
  \noalign{\smallskip}
Hubble-type& \# of galaxies & $B/T$ &  $n_{\rm b}$  &  \# of galaxies & $\left<\Delta(g-i)\right>$    &     $r_{\rm e,b}$/$r_{\rm e,d}$   \\ 
bins & for $B/T$ \& $n_{\rm b}$ &   &  & for $\left<\Delta(g-i)\right>$  \& $r_{\rm e,b}$/$r_{\rm e,d}$   & (mag)    &        \\ 
\noalign{\smallskip}
\hline
\hline
\noalign{\smallskip}       

E           & 23  &  $0.7 \pm 0.04$      & $3.5 \pm 0.5$ & 19  &  $0.03 \pm 0.04$    &   $0.3 \pm 0.2$  \\
S0--Sa  &  24 &   $0.54 \pm 0.06$  &   $1.9 \pm 0.4$ & 23 &   $0.05 \pm 0.1$   & $0.29 \pm 0.08$   \\
Sb--Sc  &  53 &   $0.2 \pm 0.04$     &  $1.8 \pm 0.3$ & 53 &   $0.28 \pm 0.06$   &  $0.24 \pm 0.08$   \\ 
Sb--Sc  &  49 &   $0.17 \pm 0.03$     &  $1.8 \pm 0.2$ & 49 &   $0.3 \pm 0.07$   &  $0.23 \pm 0.03$   \\ 
Sd--Irr   &  20  &  $0.07 \pm 0.02$  &   $0.9 \pm 0.2$  & 20 &   $0.28 \pm 0.1$   &  $0.31 \pm 0.05$    \\  
 \noalign{\smallskip}
\hline
\end{tabular}
\label{table:avegval} 
\end{table*}

\begin{figure}
\centering
\includegraphics[width=0.48\textwidth]{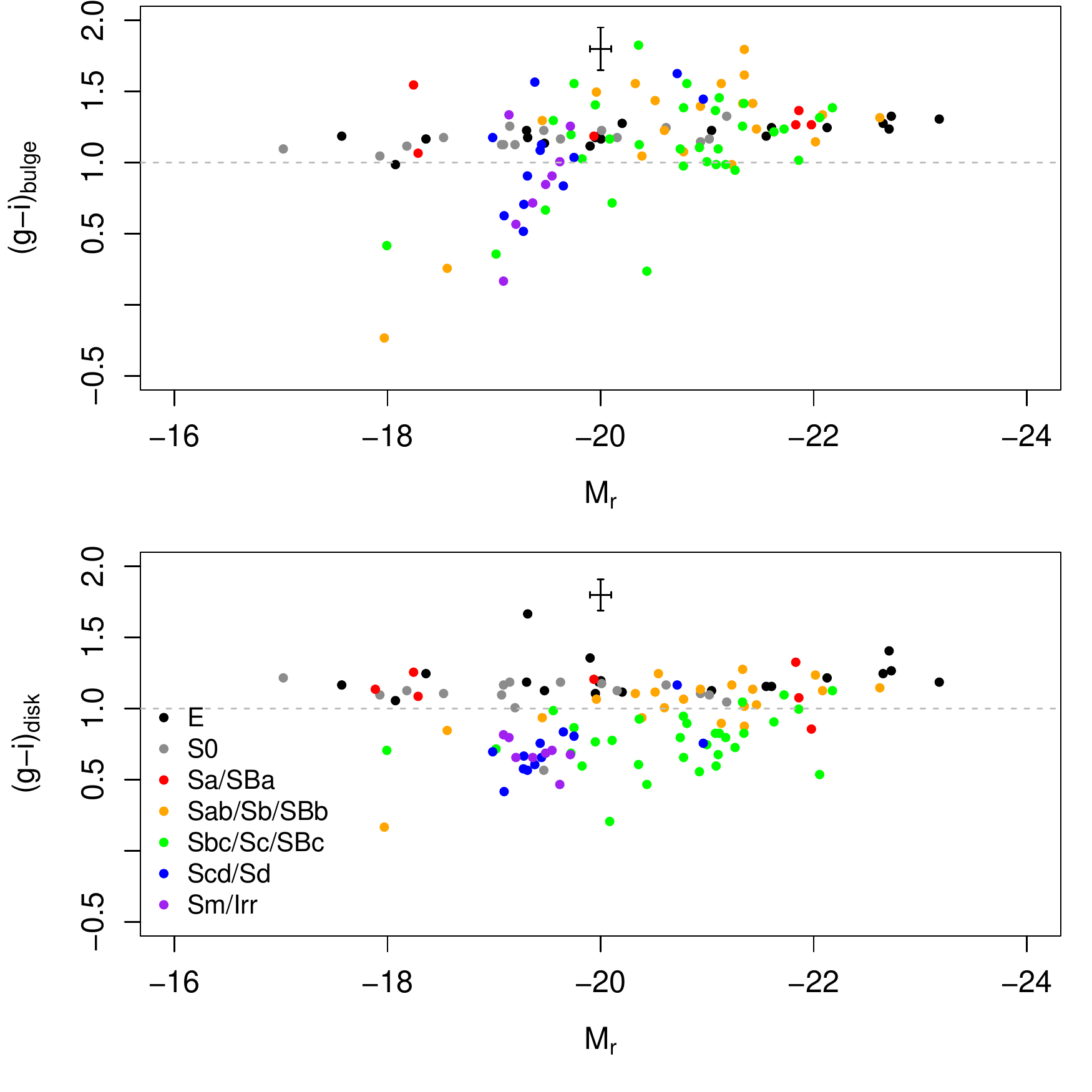}
\caption{The $g-i$ colour of the bulge (top panel) and disc (bottom panel) component as a function of $r$-band absolute magnitude. Representative error bars for our measurement are displayed in the top part of each panel. See text for further discussion on the uncertainty measurements.}
\label{fig:colbul}
\end{figure}

\subsection{Bulge-to-total}

First we consider the relative fluxes of the components. We remind the reader that magnitude is the only parameter in the \fitc fits that is entirely free to vary with wavelength. 
In Fig.~\ref{fig:BThist} we show the distribution of $r$-band bulge-to-total flux ratio, using trustworthy measurements from \fitc fits to the original images.

Only a few elliptical galaxies have $B/T$ close to one, contrary to expectations. Instead, most have $B/T \sim 0.7$, while four have $B/T < 0.6$. The four cases with small $B/T$ are NGC4636 ($B/T=0.36$), NGC4649 ($B/T=0.49$), IC3653 ($B/T=0.55$), NGC4473 ($B/T=0.58$).  
The S0--Sa galaxies display a broad distribution with a peak around $B/T=0.5$.
In the lower panel we show intermediate and late disc galaxies. Sb--Sc galaxies show an extended distribution with about three-quarters having $B/T < 0.5$, and with a noticeable peak at  $B/T < 0.1$, while all Sd-Irr galaxies have $B/T < 0.3$.

In Table~\ref{table:avegval} we give the average values of the $B/T$ for each Hubble-type bin, as measured using the original images.
The uncertainties on the median are estimated as $1.253 \sigma/\sqrt{N}$, where $\sigma$ is the standard deviation and $N$ is the number of galaxies in each Hubble-type bin.
We also measure median values when excluding galaxies with $r_{\rm e,b}>r_{\rm e,d}$. 
For these galaxies we suspect that the bulge component fits part of the disk component leading to high $B/T$ values. 
We only show this second set of results for Sb--Sc galaxies. The remaining morphological bins either contain no galaxies with $r_{\rm e,b}>r_{\rm e,d}$ or the median values do not change.

In Fig.~\ref{fig:BTz} we investigate our ability to recover the bulge-to-total flux ratio as our sample becomes more distant. For easier readability of the plot, we only show the $r$-band results for all images created with \ferengi. 
We include only those galaxies used in Fig~\ref{fig:BThist}. 
To facilitate the comparison with the results for the original images, we add the median value of the original images at redshift zero. Following the same colour coding of different Hubble classifications as in Fig.~\ref{fig:BThist}, we plot the median $B/T$ value for each group.

Both \fitb and \fitc show similar behaviour; $B/T$ decreases at high redshifts for the elliptical and early-disc samples, whereas the later-discs show the opposing trend. 
As the image resolution decreases and our ability to distinguish two components diminishes, there appears to be a tendency for the two functions to split the total flux equally.
However, the \fitc median values show smaller fluctuations and are more stable to significantly higher redshifts than \fitb. 
For instance, the late-types appear well-recovered out to $z\sim0.09$ with \fitc, while the \fitb results are only stable to $z\sim0.05$.

We can compare our average $B/T$ values (from the original images) with those found in \citet{tex:WJ09} and \citet{tex:LS10}.
 Using a sample of 143 galaxies observed in the $H$-band, \citet{tex:WJ09} found that $\sim 69$\% of bright spirals have $B/T<0.2$ and $76$\% of the bulges have $n<2$. 
We find that $67$\% of our spiral sample (Sa--Sd) have $B/T<0.2$ and $n<2$. 
Similarly, \citet{tex:LS10} used a sample of $\sim300$ S0--Sm galaxies with images in the $K$-band found to determine $B/T$ for each Hubble category: $B/T_{\rm S0} = 0.39\pm0.13$, $B/T_{\rm Sa}=0.26\pm0.12$, $B/T_{\rm Sb}=0.12\pm0.09$ and $B/T_{\rm Sd}=0.06\pm0.13$.
Despite some differences in our sample selections and the wavelength considered, our $B/T$ distributions for different morphologies are highly consistent with these two independent studies.

\subsection{Colours}


Galaxies display a range of optical colours, which correlate strongly with morphology and structure (e.g., see \citealt{tex:WL13}).  
However, the total colour of a galaxy averages over any distinct stellar populations it may contain. 
The bulge and disc components of a galaxy may be expected to comprise contrasting stellar populations due to their different formation histories. Considering their colours individually is thus a sensible first step toward better understanding the distribution of stellar populations within galaxies.
For example, quantifying the differences and correlations between the colours of bulge and disc components can help us differentiate between proposed bulge formation scenarios.

Previous work has suggested that there are substantial variations in the colours of bulges and disks between galaxies, while the colours of the two components within a given galaxy are often similar \citep{tex:PB96}, though significantly offset \citep{tex:MC04,tex:CD09}.
Here we briefly present the results for our initial sample, and show the advantage of the multi-band technique method for measuring bulge and disc colours. 

In Fig.~\ref{fig:coldiffm} we plot the distribution of the colour difference between the two components. In the top panel we see that early-type galaxies contain bulges and disks with similar colours.
In contrast, the late-types possess bulges that are significantly redder than their discs. The average values of the component colour difference can be found in Table~\ref{table:avegval}.
For all spiral galaxies (Sa--Sm), we find $\left<\Delta(g-i)\right>\sim 0.3$~mag, irrespective of their more detailed Hubble type.
Even S0s, considered alone, typically possess bulges that are slightly redder than their discs, with $\left<\Delta(g-i)\right> = 0.05\pm0.05$~mag.  

These values compare very well with previous measurements in the literature.
\citet{tex:MC04} found an average bulge$-$disk colour difference of $\left<\Delta(B - R)\right> = 0.29 \pm 0.17$~mag for a sample of 172 low-inclination disc galaxies (S0--Irr), while \citet{tex:HS10} find $\left<\Delta(B - R)\right> = 0.23 \pm 0.02$~mag for $L_\ast$ discs in eight low-redshift clusters.
Similarly, \citet{tex:CD09} reported a colour difference of $\left<\Delta(u-r)\right>= 0.27\pm0.04$~mag (without their average dust correction) using $\sim1500$ two-component galaxies extracted from the Millennium Galaxy Catalogue.
The bulge$-$disc colour difference we find for S0s is also consistent with the $\left<\Delta(g-i)\right> = 0.09\pm0.01$~mag found by \citet{tex:HL14} for S0s in Coma.

To examine this behaviour in more detail, Fig.~\ref{fig:colbul} presents the colour-magnitude distribution for each component, colour-coded by Hubble-type. 
Elliptical and S0 galaxies typically have both their components on a red-sequence, with $(g-i) \sim 1.2$~mag, resulting in the distributions centred around zero in the top panels of Fig.~\ref{fig:coldiffm}.  The disc colours of early-spirals (Sa/Sab) are also typically on this red-sequence, while the discs of late-spirals (Sb--Sm) inhabit a blue cloud, with later types being fainter (though our heterogeneous sample selection may be somewhat responsible for this).  The bulge colours of spirals show a considerable scatter.  Some, particularly those of types Scd--Sm, lie in the blue cloud, whereas the bulges of Sab--Sc galaxies are often above than the red-sequence.  Dust extinction may be responsible for these very red bulges.  However, we do not see any significant trend in bulge colour with disc inclination, as one might expect if this were the case.

We now consider te behaviour of the bulge$-$disc colour difference versus apparent redshift. In the top panel of Fig.~\ref{fig:coldiff}, we show the \fitc results. The early-type galaxies have a median colour difference very close to zero, which remains almost constant out to $z \sim 0.15$. For Sb--Sc galaxies the median value is stable till redshift 0.03, after which it is overestimated with respect to the original measurement, but at least relatively stable and differentiated from the early-types. Sd--Irr galaxies show a greater degree of variation beyond $z \ga 0.05$, although note that this sample contains intrinsically fainter galaxies than the other sets.

In the middle panel of Fig.~\ref{fig:coldiff} we show the results of fitting using the aperture method, for which structural parameters are fixed to the $r$-band results and colours obtained by fitting for the bulge and disc fluxes in the each other band.
The initial behaviour is similar to that in the top panel, but with greater variation, such that the different Hubble types are less clearly differentiated beyond $z \ga 0.06$. However, we notice that \fitc median colours for the Sb- Sc galaxies beyond redshift 0.08 are maintained to higher values, compared to the colour at redshift zero, while the aperture median colour drops again close to the dashed line.
 
Finally, in the bottom panel, we show the \fitb results, from independent fits to each band.  In this case the variations in structural parameters between bands make the colours of each component very noisy, and sensible values cannot be obtained beyond $z \ga 0.03$.  This emphasises that colours for the bulges and discs of individual galaxies cannot be directly obtained via independent fits to multiple bands.  Even using such measurements in a statistical fashion (e.g. to estimate the average colours of bulges) would be highly unreliable.

\begin{figure}
\centering
\includegraphics[width=0.48\textwidth]{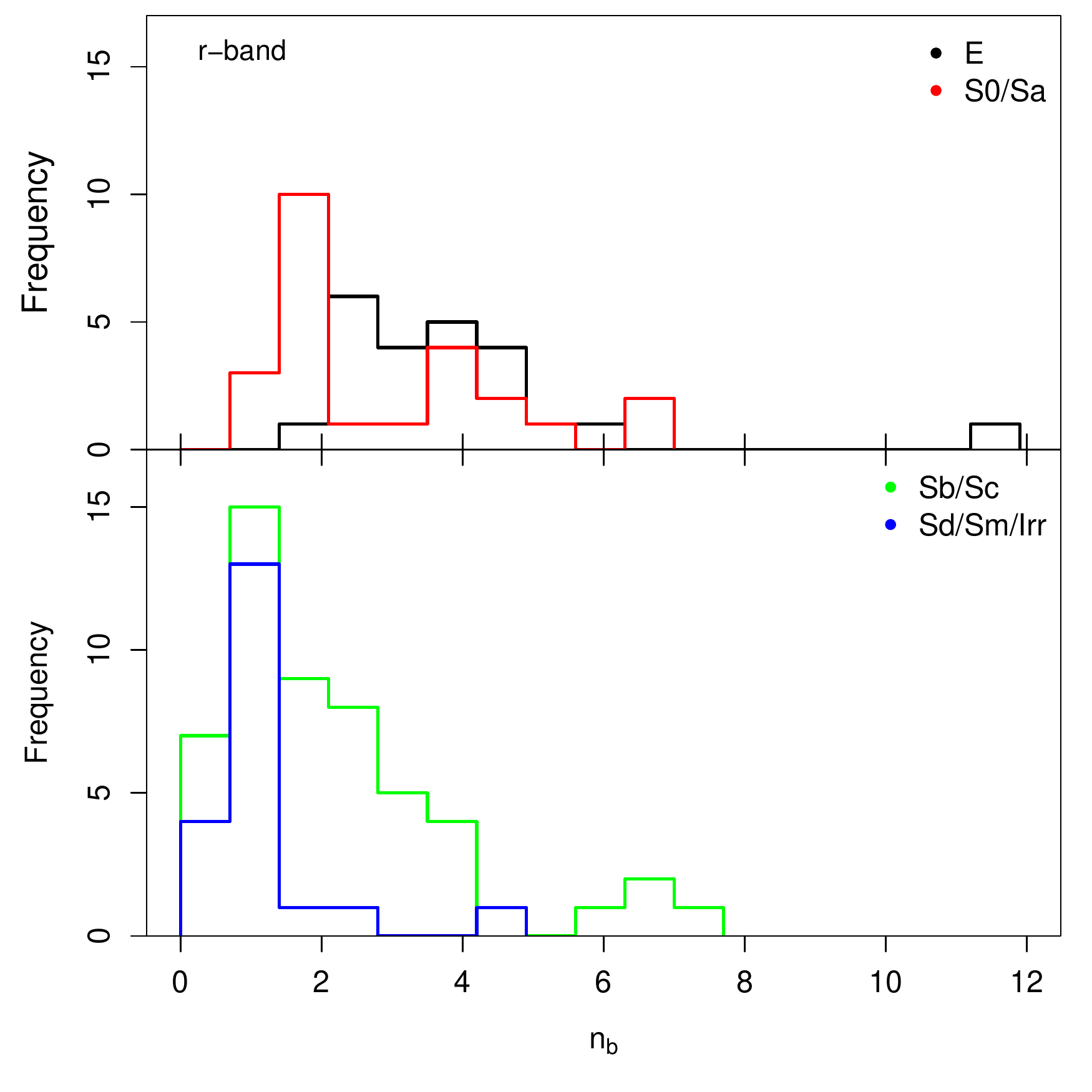}
\caption{The distribution of bulge \sersic index for the fits to the original images. Only galaxies with a significant bulge are shown in this figure. A more detailed presentation of the bulge \sersic index distribution, particularly for small values  ($n_{\rm b}<1$), can be found in Fig. \ref{fig:BTnb}.
}
\label{fig:n_montage}
\end{figure}

\begin{figure}
\centering
\includegraphics[width=0.48\textwidth]{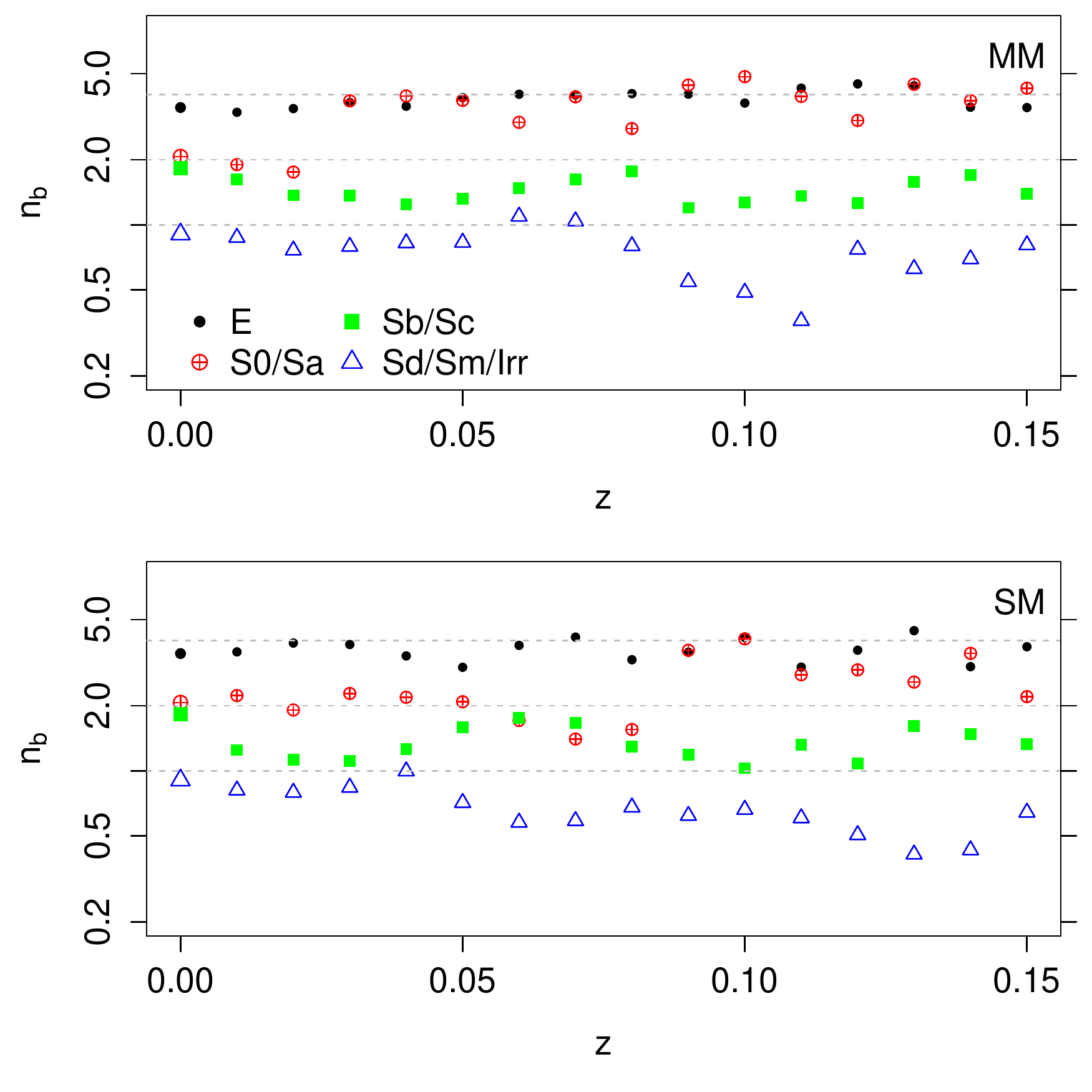}
\caption{The median value of the bulge \sersic index distribution, as seen in Fig.~\ref{fig:n_montage}, as a function of apparent redshift. At redshift zero the results from the original images are plotted, while for higher redshifts we use the results for the artificially-redshifted images. Horizontal lines are plotted at $n_{\rm b}=1$, $2$ and $4$.}
\label{fig:n_ferengi}
\end{figure}

\subsection{\sersic index}
\label{sec:sindex}

We now move on to study how the structural parameters ($n_{\rm b}$, $r_{\rm e,b}$, $r_{\rm e,d}$) are distributed for different Hubble types, and investigate the performance of the multi-band fitting in measuring these values.
In Fig.~\ref{fig:n_montage} we plot the distribution of the bulge \sersic index, as measured in the original images using the \fitc method. 
Elliptical galaxies present a peak around $4$.  S0--Sa galaxies display a bimodality, with peaks around $1$--$2$ and $4$.
Intermediate spiral types (Sb--Sc) have a wide range of $n_{\rm b}$ values, mostly in the range $\sim 1$--$4$. 
The five Sb--Sc galaxies with $n_{\rm b}>4$ are NGC5430, NGC2841, NGC3521, NGC3642 and NGC4698, with $n_{\rm b} = 4.2$, $5.7$, $6.7$, $6.8$ and $7.1$, respectively. See Appendix~\ref{sec:notes} for further discussion of some of these galaxies.
All our Sd--Irr galaxies have bulges with $n_{\rm b} < 2$, except for NGC4653 and NGC4108B, with $n_{\rm b} = 2.7$ and $4.3$.
The average values are given in Table~\ref{table:avegval}.  Typical spiral bulges with $n_{\rm b} \sim 2$ and a progression to lower bulge \sersic index for later Hubble types corresponds very well to expectations from the literature (e.g., \citealt{tex:GW08,tex:LS10,tex:MC11,tex:HL14}).

Figure~\ref{fig:n_ferengi} uses our artificially-redshifted images to examine how well we are able to recover the bulge \sersic index with increasing redshift.  This plot is complicated by the bimodal distribution of the S0--Sa class, which results in the median being unstable.  Overall, both \fitc and \fitb methods recover similar median $n_{\rm b}$ values.   For the elliptical galaxies the median $n_{\rm b}$ is well recovered at all redshifts probed.  For the spiral classes, the median $n_{\rm b}$ is more variable, particularly for $z \ga 0.05$.  However, in general, the \fitc fits appear to be rather more stable.

\subsection{Effective radius}

\label{sec:re}

\begin{figure}
\centering
\includegraphics[width=0.48\textwidth]{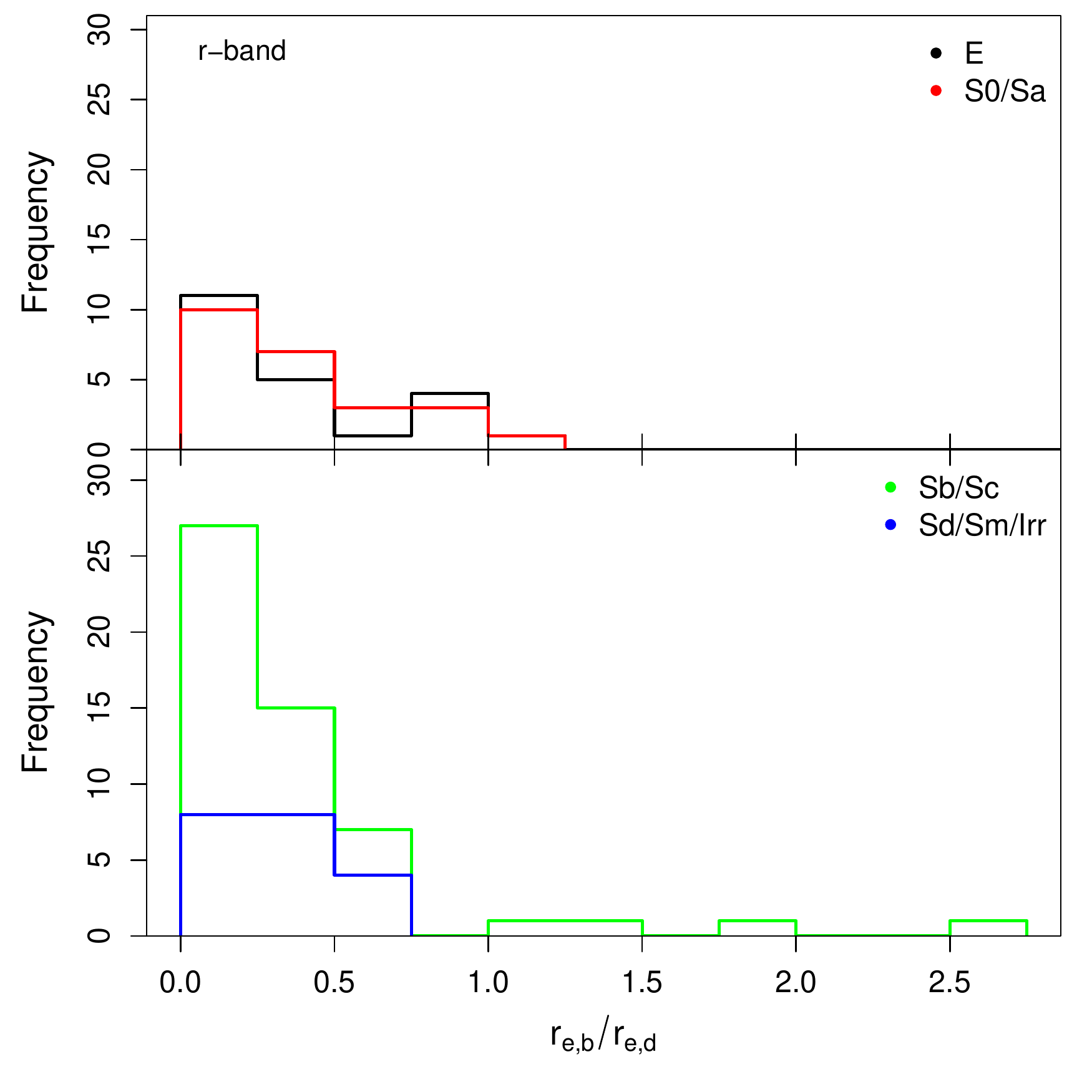}
\caption{The distribution of the ratio of bulge and disc effective radii, as measured from the original images. 
Only galaxies with both a significant bulge and disc are shown in this figure. In the top panel there is one galaxy, NGC4458, outside the axes, with $r_{\rm e,b}$/$r_{\rm e,d} = 5$. 
An alternative presentation of the $r_{\rm e,b}/r_{\rm e,d}$ distribution can be found in Fig. \ref{fig:BTrratio}.} 
\label{fig:r_montage}
\end{figure}

In Fig.~\ref{fig:r_montage} we investigate the relationship between the sizes of the bulge and disc and Hubble type, by plotting histograms of the ratio of bulge effective radius to disc effective radius. Note that we do not constrain our bulges to be smaller than our discs, and neither do we subsequently exclude galaxies based on  $r_{\rm e,b}/r_{\rm e,d}$. Consequently, in Fig~\ref{fig:r_montage} we find seven galaxies\footnote{black: IC3653 and NGC4458; red: NGC4452; green:  NGC3521, NGC3642, NGC3893 and NGC4698.} with $r_{\rm e,b}/r_{\rm e,d}>1$.  Most of these galaxies have peculiarities that interfere with the fit.  They are discussed individually in Appendix~\ref{sec:notes}.

Disregarding the few galaxies with $r_{\rm e,b}/r_{\rm e,d}>1$, we find very little difference between the distributions for different Hubble types.  Average values are listed in Table~\ref{table:avegval}.   Other studies also tend to find little dependence of the bulge-to-disc size ratio on morphology (e.g., \citealt{tex:GW08}).

We find that bulges are typically around one-quarter of the size of their accompanying discs. Rather than the ratio of effective radii, other studies typically quote $r_{\rm e,b}/h$, where $h=r_{\rm e,d}/1.678$ is the exponential disc scalelength.  Furthermore, $h$ is often corrected for inclination-dependent projection and extinction effects, which complicates comparisons.  Finally, given the strong wavelength dependence of galaxy effective radius found by \citet{tex:VB14}, measurements at optical versus near-infrared wavelengths may be expected to differ significantly, even when an average extinction correction is applied.  Nevertheless, we attempt an approximate comparison.

Assuming some average corrections, our median optical $r_{\rm e,b}/r_{\rm e,d} \approx 0.25$ translates into an extinction-corrected, face-on $r_{\rm e,b}/h \approx 0.35$.  This agrees well with the values found by \citet{tex:KW00}, \citet{tex:Nv07} and \citet{tex:MA08}, but is a factor of $1.5$ larger than found by the careful analysis of multiple datasets from the literature \citet{tex:GW08} and twice that found by \citet{tex:LS10}.  The latter study, and some of the works that were included in \citet{tex:GW08}, included additional central components in their models, such as bars or nuclei.  This may have led to the smaller bulge sizes they measure.  Given the care taken in these studies, we suspect that our bulge $r_{\rm e,b}$ may be somewhat overestimated.  However, remember that our aim in this work is to perform fits to our nearby galaxies in a simple, automated manner, suitable for large surveys of relatively distant galaxies, and ascertain the performance of this approach.

\begin{figure}
\centering
\includegraphics[width=0.48\textwidth]{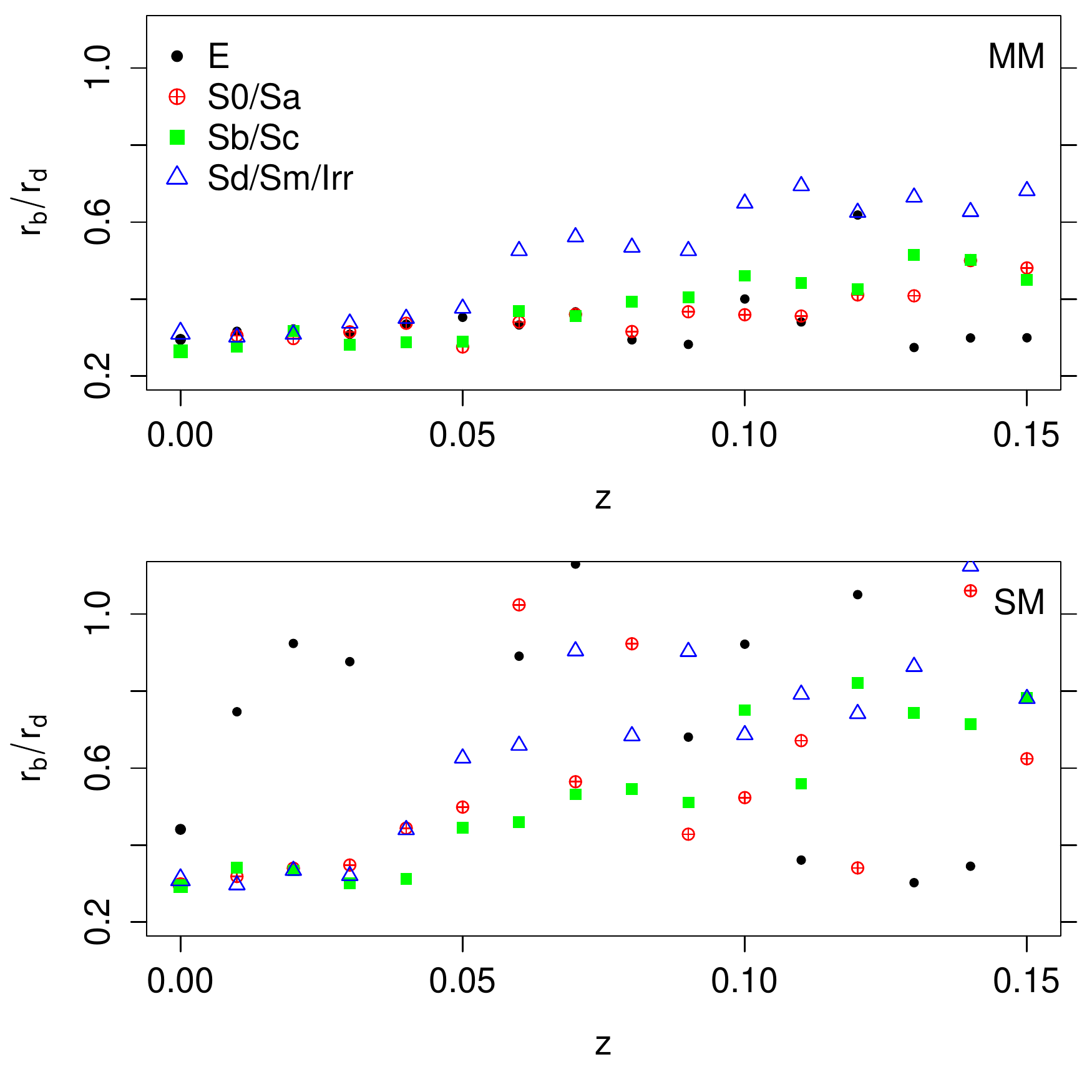}
\caption{The median value of the $r_{\rm e,b}/r_{\rm e,d}$ distribution, as seen in Fig.~\ref{fig:r_montage}, as a function of apparent redshift. 
At redshift zero the results from the original images are plotted, while for higher redshifts we use the results for galaxies fit in the artificially-redshifted images.}
\label{fig:r_ferengi}
\end{figure}

With this in mind, Fig.~\ref{fig:r_ferengi} shows the median $r_{\rm e,b}/r_{\rm e,d}$ for several Hubble type bins as a function of simulated redshift.  For the multi-band (\fitc) fits, we again see that for low redshift data there is little difference with morphology.  The average size ratios remain fairly constant to $z \sim 0.04$.  After this, as the data quality becomes substantially poorer, a strong trend to increasing  $r_{\rm e,b}/r_{\rm e,d}$ is seen, particularly affecting galaxies with lower $B/T$.  Single-band fits perform reasonably similarly (neglecting the ellipticals, for which the reality of the disc is unclear).  However, they show a stronger and noisier bias, which sets in at slightly lower redshifts.

Generally, we observe that the lower the data quality, the harder it is to separate the two components and the more similar their properties become. 
However, it is usually the bulge fit which is most affected, and hence biased.  The effective radius of the disc components tend to remain stable for almost the entire redshift range considered, particularly for our multi-band fits.

\section{Correlation of structure parameters}
\label{sec:correlations}

\subsection{Bulge-to-Total}

For disk galaxies, the overall \sersic index (of a single-\sersic model fit) is often regarded as an indication of the bulge-to-total ratio ($B/T$).  Indeed, we adopt this interpretation in \citet{tex:HB13,tex:VB13b,tex:VB14}.
With our bulge-disk decompositions, we are now in a position to test this. 

\begin{figure}
\centering
\includegraphics[width=0.45\textwidth]{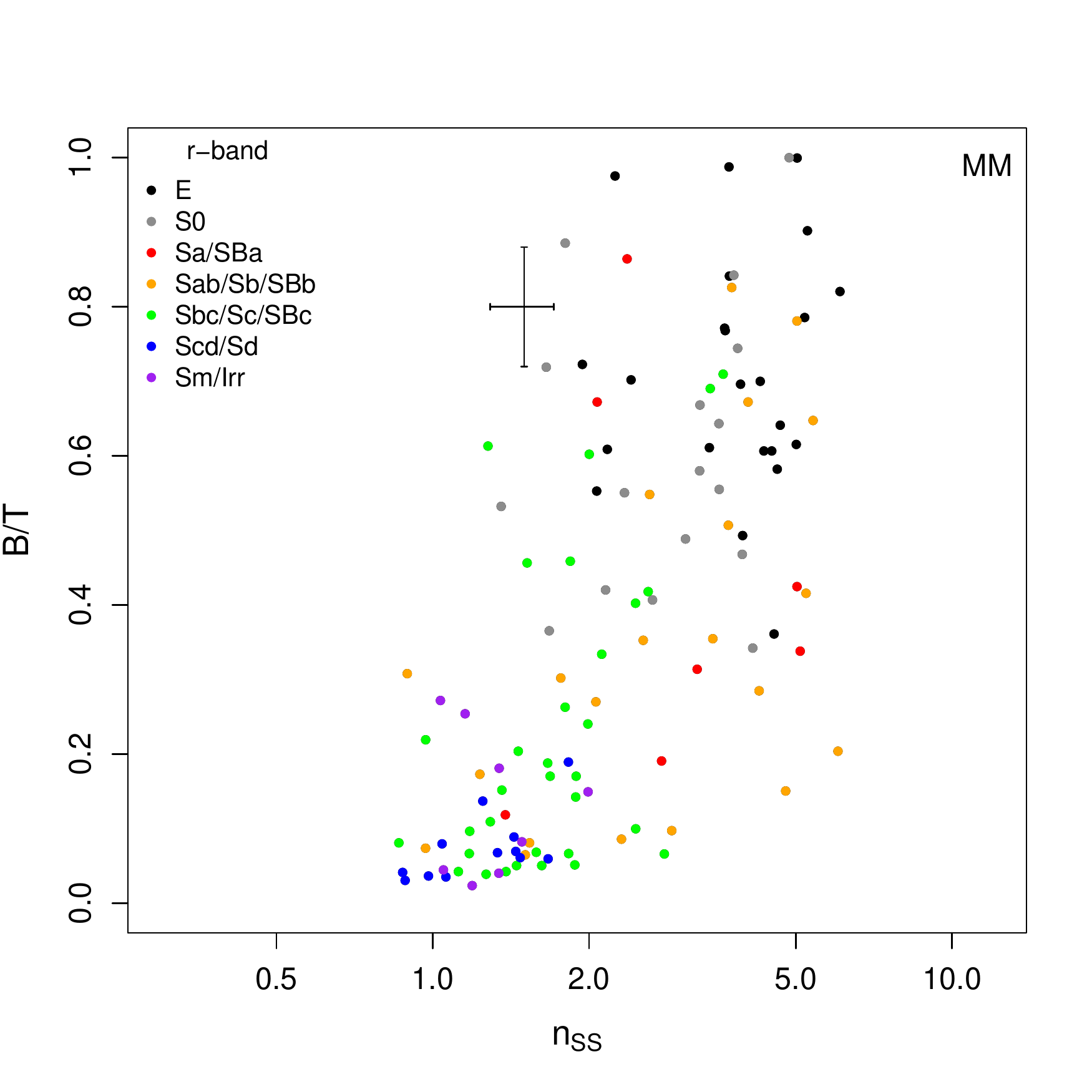}
\caption{Bulge-to-total flux ratio as a function of \sersic index for our \fitc method. Only galaxies with significant bulge measurements are shown in this figure. Representative error bars for our measurement are displayed. See text for further discussion on the uncertainty measurements.
}
\label{fig:BTn}
\end{figure}

In Fig.~\ref{fig:BTn} we plot the bulge-to-total ratio as a function of the single-\sersic index ($n_{\rm SS}$) measured in V13. 
We see that, as the overall \sersic index increases, the bulge is responsible for a greater proportion of the galaxy flux, confirming our expectations.  Galaxies with a low overall \sersic index typically contain two components, a bulge and a disc, and the more dominant the bulge component, the higher the overall \sersic index.

Galaxies of type Scd and later generally have low $B/T$, while earlier spirals (Sa--Sc) span a wide range of $B/T$.  
Interestingly, earlier types tend to have greater $n_{\rm SS}$ at a given $B/T$, suggesting that $n_{\rm SS}$ is also dependent on other aspects of galaxy structure.
Most galaxies with $n_{\rm SS}\sim1$ have $B/T<0.1$, with the exception of four galaxies\footnote{orange: NGC5624, green: NGC1084, purple: NGC0428 and NGC0853}.

\begin{figure}
\centering
\includegraphics[width=0.48\textwidth]{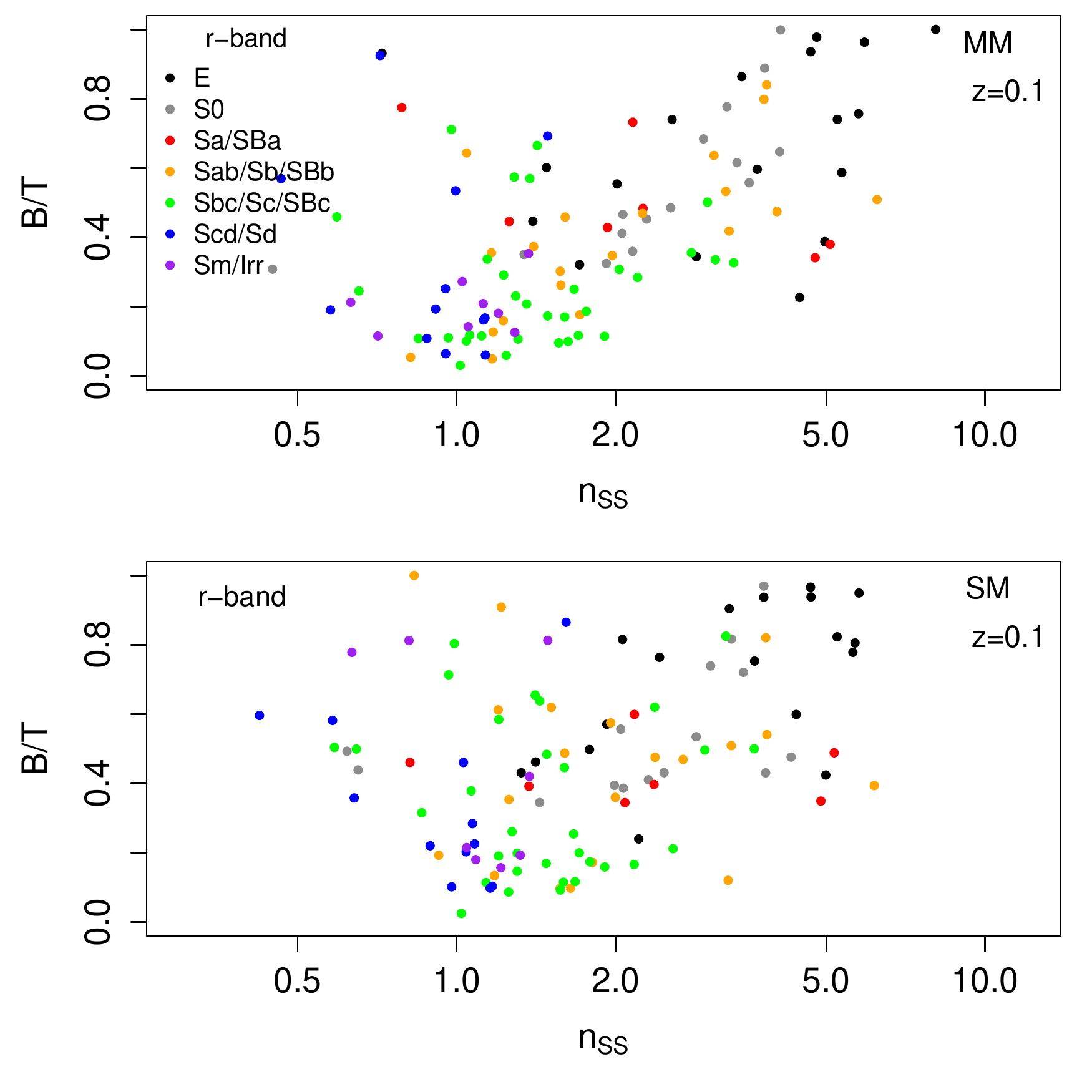}
\caption{Bulge-to-total flux ratio as a function of \sersic index for our \fitc fits to the artificially-redshifted images. Only galaxies with significant bulge measurements are shown in this figure. }
\label{fig:BTn_z}
\end{figure}

For the original images, the resolution and signal-to-noise is sufficiently good that fitting bands individually is comparable to our multi-band approach.  However, for more distant galaxies our multi-band method gives more robust measurements of $B/T$.  We demonstrate this in Fig.~\ref{fig:BTn_z}, where we again plot the bulge-to-total ratio as a function of the single-component \sersic index ($n_{\rm SS}$), but now using galaxies artificially redshifted to $z=0.1$. This figure shows the results of fitting each band independently (\fitb), in addition to our multi-band measurements (\fitc). The clearer correlation for \fitc clearly illustrates the advantage of our multi-band method.

\begin{figure}
\centering
\includegraphics[width=0.45\textwidth]{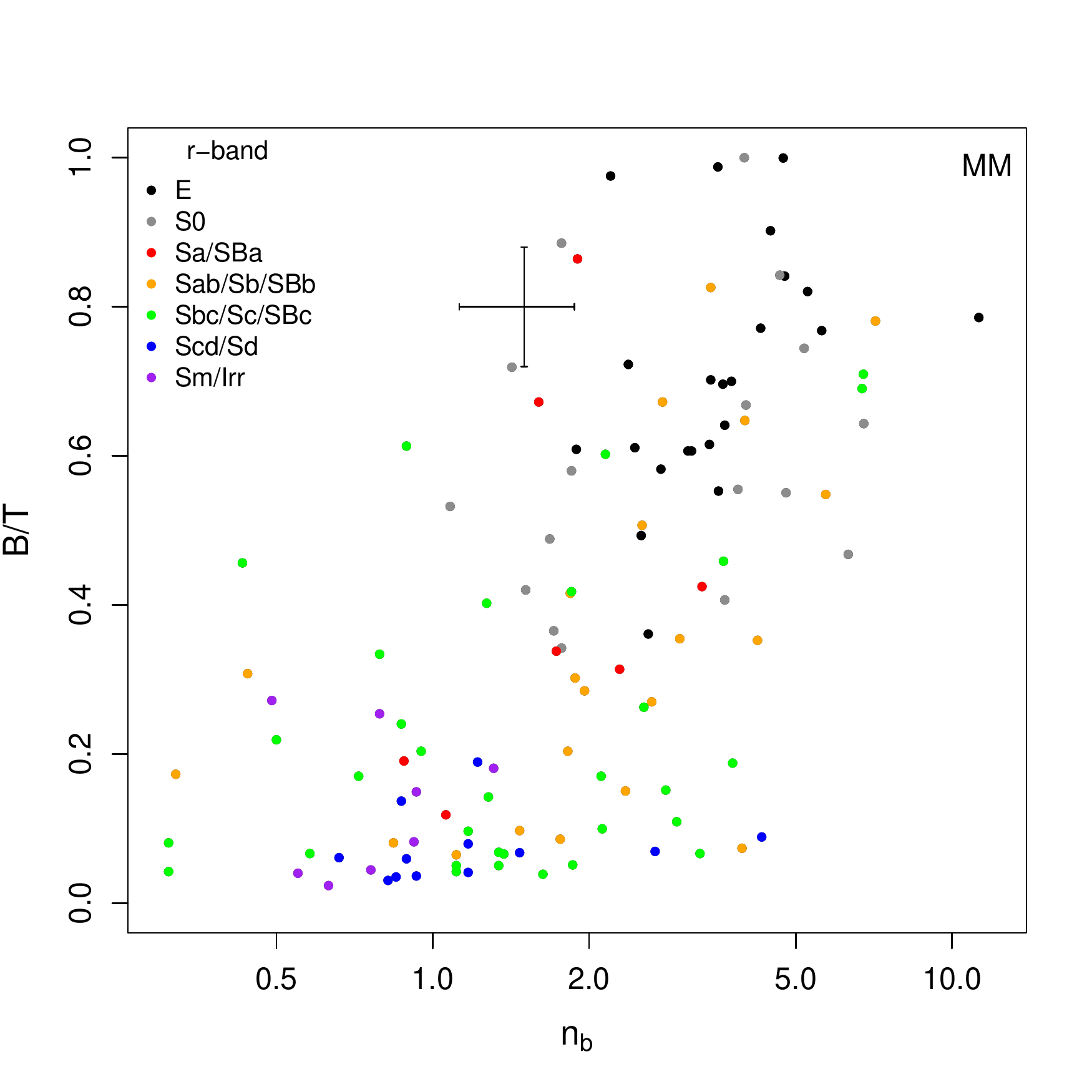}
\caption{Bulge-to-total flux ratio as a function of the bulge \sersic index from our \fitc fits. Only galaxies with significant bulge measurements are shown in this figure. Representative error bars for our measurement are displayed. See text for further discussion on the uncertainty measurements}
\label{fig:BTnb}
\end{figure}

A number of studies have presented a correlation between bulge-to-total ratio and the \sersic index of the bulge (e.g., \citealt{tex:G01}), particularly as a diagnostic for distinguishing so-called pseudo- and classical-bulges \citep{tex:DS08,tex:WJ09,tex:LS10}. Figure~\ref{fig:BTnb} confirms this relationship for our measurements.  The scatter is relatively large, probably as a result of the difficulty in constraining the bulge properties, as discussed in the previous section.  Nevertheless, it is clear that the more bulge-dominated a galaxy is, the higher its bulge \sersic index.

\begin{figure}
\centering
\includegraphics[width=0.45\textwidth]{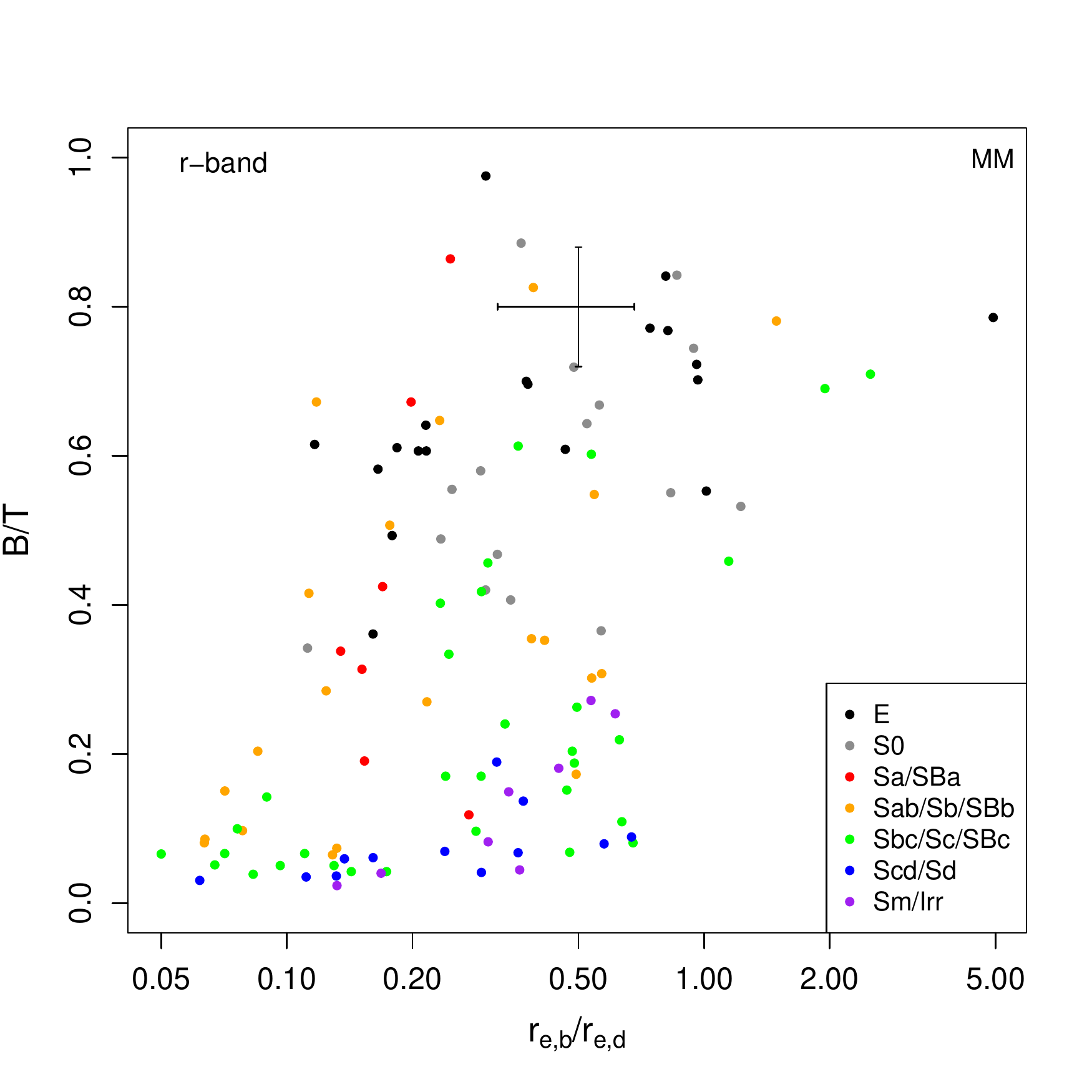}
\caption{Bulge-to-total flux ratio as a function of the bulge-to-disc size ratio. Only galaxies with both a significant bulge and disc are shown in this figure. Representative error bars for our measurement are displayed. See text for further discussion on the uncertainty measurements}
\label{fig:BTrratio}
\end{figure}

We also see a weak correlation between the ratio of bulge and disc sizes, $r_{\rm e,b}/r_{\rm e,d}$, and the bulge-to-total flux ratio, in Fig.~\ref{fig:BTrratio}.  There are some indications that the relation depends on morphological type, but the scatter and incompleteness of our sample prevent us from making definitive conclusions.

\subsection{Component axis ratios and the division of ellipticals and lenticulars}

\begin{figure}
\centering
\includegraphics[width=0.48\textwidth]{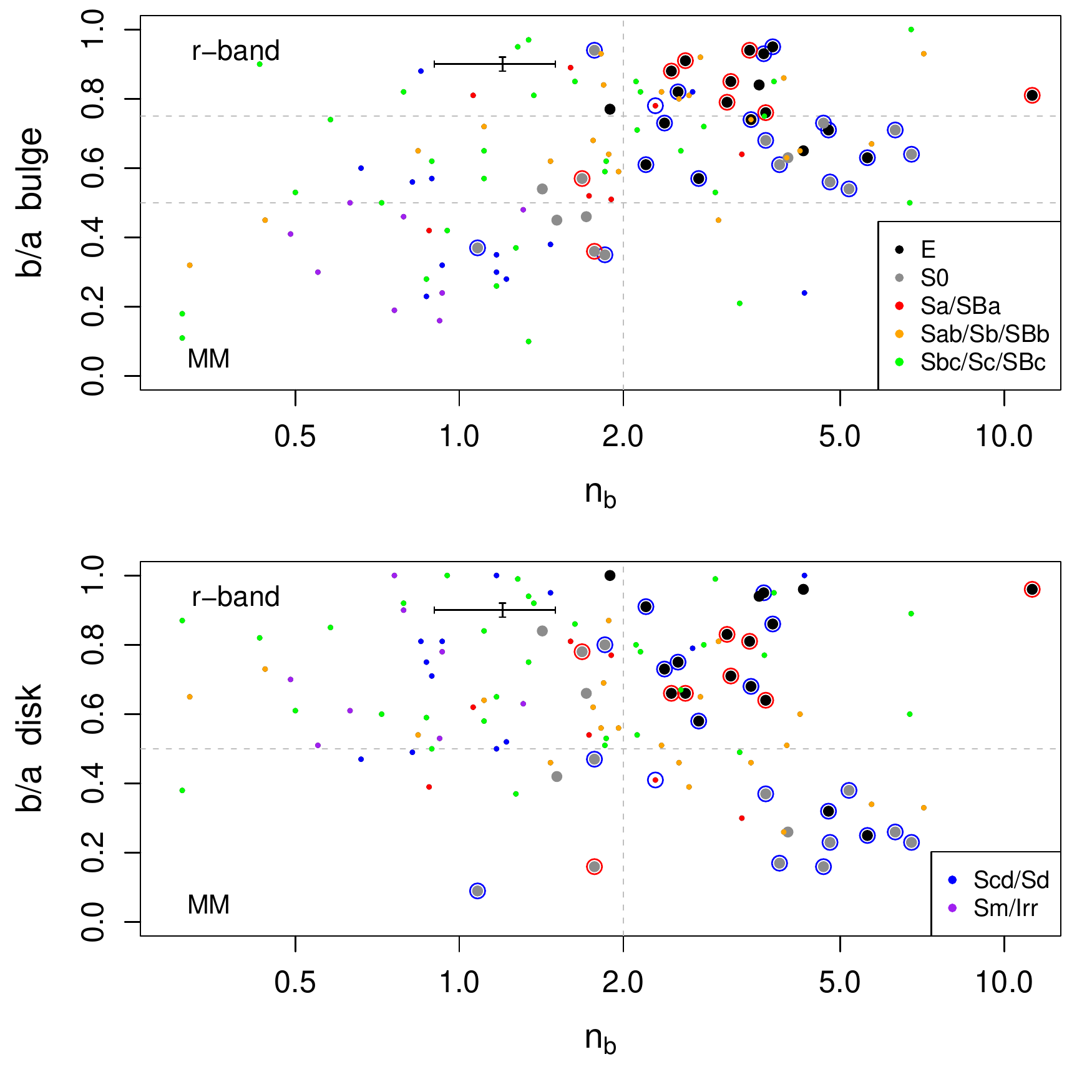}
\caption{The axis ratio of bulge (upper panel) and disc (lower panel) as a function of the bulge \sersic index. The red and blue circles indicate slow and fast rotators, respectively, as measured in \citet{tex:EC11}. Only galaxies with both a significant bulge and disc are shown. Representative error bars for our measurement are displayed in the top part of each panel. See text for further discussion on the uncertainty measurements.
}
\label{fig:axisratio}
\end{figure}

Figure~\ref{fig:axisratio} explores the relationship between the axis ratio of the bulge (top panel) and disc (bottom panel) versus the bulge \sersic index.  As before, we include galaxies with elliptical morphologies in these plots for two reasons.  First, recent work has blurred the lines between elliptical and lenticular galaxies, with many ellipticals found to contain faint disc components when studied in detail \citep{tex:KA13}.  Second, our aim is to inform work on large surveys, which may not have morphological classifications available.  For these ellipticals, and despite our nomenclature, we do not go so far as to assume that the exponential component of our model represents a disc, but consider it to be an indication of additional structure that cannot be well-modelled by a single \sersic component.

Considering all the points in the top panel of Fig.~\ref{fig:axisratio}, there is a clear correlation such that bulges with higher \sersic index tend to be more circular ($b/a \sim 1$).  The vast majority of galaxies with elliptical morphologies are found in the upper-right region, with $n_{\rm b} > 2$ and $b/a > 0.5$, as might be expected.  Lenticulars slightly separate out from ellipticals in this plot, generally being limited to slightly lower $b/a$ and a wider range of $n_{\rm b} > 2$.

Moving our attention to the lower panel, we first see little correlation between disk axial ratio and bulge \sersic index.  Note that our sample of spiral galaxies is seriously incomplete for inclined systems, due to selection restrictions applied by the studies from which V13 obtained their sample.

Focussing on ellipticals and S0s, we see a surprisingly strong separation between the two morphologies in disc $b/a$.  
The vast majority of galaxies with classified as elliptical have $(b/a)_{\rm disc} > 0.5$, while the lenticulars have mostly $(b/a)_{\rm disc} < 0.5$.  
We also see an offset of the inclined lenticulars to higher bulge \sersic indices.  We are not certain whether this reflects reality, or is a bias in the decomposition process.  Simulations suggest that only small variations in $n_{\rm b}$ are expected from decomposition effects \citep{tex:PP13b}.  
S0s are generally not expected to contain significant amounts of dust, so extinction should not play a significant role.  In any case the effects are typically $\la 0.1$ in $n_{\rm b}$.

In order to explore the separation of ellipticals and lenticulars in Fig.~\ref{fig:axisratio} further, we highlight early-type galaxies using their kinematic classification from \citet{tex:EC11}.  
We note that almost all the early-type galaxies with low $(b/a)_{\rm disc}$ are fast rotators, including both of the elliptical galaxies which fall in this region of parameter space dominated by S0s.
The early-types with $(b/a)_{\rm disc} > 0.5$ are a mixture of fast and slow rotators, however most (or all with $n_{\rm b} > 2$) have been classified as elliptical galaxies (see \citealt{tex:CE11b} and \citealt{tex:KA13} for a more thorough study of this topic).

Obviously a plot of $(b/a)_{\rm disc}$ versus $n_{\rm b}$, or even just $(b/a)_{\rm disc}$ alone, is an effective, automated way of separating galaxies with elliptical and lenticular visual morphologies.  However, this raises the question of whether such a separation is a physically meaningful thing to do.

The difficulty of distinguishing face-on S0s from ellipticals is a well known problem.  The result in Fig.~\ref{fig:axisratio} illustrates this issue in terms of quantitative structural measurements.  When a fast-rotating early-type galaxy appears to have an inclined disc ($(b/a)_{\rm disc} \la 0.5$, it is usually classified as S0. If the same galaxy were viewed closer to face-on, it would be classified as elliptical.  In our (non-representative) sample, this amounts to about half of true S0 galaxies (discy, fast-rotators that that would have been visually classified as S0 if viewable from other angles) being misclassified as elliptical.  We presently do not have a reliable morphological or structural way of recovering these objects, but instead must resort to kinematic information (e.g., \citealt{tex:KA13}).  However, we remain hopeful that with additional work we can make further progress on an image-based solution to this long-standing problem.

\section{Conclusions}  
\label{sec:conclusions}

All previous studies that have utilised one- or two-dimensional photometric bulge-disc decompositions have performed their fits using, at most, two bands simultaneously.
In this work, for the first time, we have performed bulge-disc decompositions simultaneously on five-band imaging.
To achieve this, we have used \galfitm, a modified version of \galfitthree which enables a single, wavelength-dependent model to be fit to multiple images of the same galaxy.

We have evaluated the performance of our multi-band method by applying it to SDSS $ugriz$ images of 163 nearby galaxies, as well as to another 3863 artificially-redshifted images of the same objects.
For our models, we use a combination of two \sersic profiles: one for the disc, with $n$ fixed to one, and another for the bulge, with free-$n$.

Using the original images, we have shown that our fitting results generally agree well with structural parameters obtained from the literature, both when we compare specific galaxies and averages for bins of Hubble type.
We confirm the standard picture that disc galaxies of earlier morphology have a larger fraction of their flux in a bulge component.
However, the sizes of these bulges, with respect to their accompanying disc, do not vary significantly with Hubble type.
We find that bulges in spiral galaxies typically have a \sersic index $n \sim 2$, except in the latest-types, where it is more usually exponential $n \sim 1$.
The dominant component of ellipticals has $n \sim 3.5$.
Puzzlingly early-type discs appear to display a bimodality with peaks around $n \sim 1$ and $4$.

The observed colours of disks display a classic colour-magnitude diagram, with a well defined red sequence, inhabited by E-Sb galaxies, and a blue cloud corresponding to later Hubble types.
The colour-magnitude diagram for bulges is more complex.
The bulges of E-Sa galaxies lie on a red-sequence similar to their disks, with only a small average difference in the colours of their bulges and discs.
The bulges of Sb-Sc galaxies are often even redder than the early-type red-sequence, indicative of dust reddening.
On the other hand, for many late-type disc galaxies we find bulges with fairly blue colours, suggestive of recent star-formation.
Despite this complexity, the average difference in the colours of bulges and discs within the same galaxy is constant for all spiral galaxies, $\left<\Delta(g-i)\right> \sim 0.3$~mag.

Our fits permit a quantitative illustration of the notorious difficulty of distinguishing between galaxies with elliptical and lenticular morphology.
Most early-types are well-fitted by a combination of a \sersic profile and an exponential profile.
It is not clear how often this exponential profile represents a genuine disc component, or whether it reveals the presence of an extended halo or some other substructure.
Nevertheless, a significant inclined disc component is strong indication that an early-type galaxy will be visually classified as S0 and possess fast-rotator kinematics.
Unfortunately, the lack of S0 morphologies among galaxies with face-on disc components, despite kinematic evidence indicating the discs are real, suggests that such systems are usually misclassified as ellipticals.

Using our artificially-redshifted images we have investigated the range of redshift over which our fit parameters remain stable, and hence are inferred to be reliable.
We have demonstrated that the results of our multi-band fits show less variation and are reliable to higher redshifts than the results of fitting each band independently.
This is true even in the common single-band approach where one first performs a full fit on a preferred band, then fixes the structural parameters in all subsequent fits to those obtained in that preferred band.

Our approach produces somewhat more reliable measurements of the bulge \sersic index and effective radius, although these are both difficult quantities to measure with consistent accuracy, especially when it is not feasible to fit each galaxy interactively.
More promising is our method's performance in recovering the bulge-to-total flux ratio and in differentiating between the colours of the bulge and disc.
We therefore recommend the adoption of multi-band bulge-disc decomposition, allowing studies to reliably probe to greater distances and lower-luminosity galaxies.

\section*{Acknowledgments}
This publication was made possible by NPRP grant \# 08-643-1-112 from the Qatar National Research Fund (a member of Qatar Foundation). The statements made herein are solely the responsibility of the authors.  BH and MV were supported by the NPRP grant.  SPB gratefully acknowledges an STFC Advanced Fellowship.  We would like to thank Carnegie Mellon University in Qatar and The University of Nottingham for their hospitality.  We would like to thank the referee for the constructive comments that improved the paper.

Funding for the SDSS and SDSS-II has been provided by the Alfred P. Sloan Foundation, the Participating Institutions, the National Science Foundation, the U.S. Department of Energy, the National Aeronautics and Space Administration, the Japanese Monbukagakusho, the Max Planck Society, and the Higher Education Funding Council for England. The SDSS Web Site is http://www.sdss.org/. The SDSS is managed by the Astrophysical Research Consortium for the Participating Institutions. The Participating Institutions are the American Museum of Natural History, Astrophysical Institute Potsdam, University of Basel, University of Cambridge, Case Western Reserve University, University of Chicago, Drexel University, Fermilab, the Institute for Advanced Study, the Japan Participation Group, Johns Hopkins University, the Joint Institute for Nuclear Astrophysics, the Kavli Institute for Particle Astrophysics and Cosmology, the Korean Scientist Group, the Chinese Academy of Sciences (LAMOST), Los Alamos National Laboratory, the Max-Planck-Institute for Astronomy (MPIA), the Max-Planck-Institute for Astrophysics (MPA), New Mexico State University, Ohio State University, University of Pittsburgh, University of Portsmouth, Princeton University, the United States Naval Observatory, and the University of Washington.

\bibliography{references}

\appendix

\begin{figure*}
\centering
\includegraphics[height=3.2cm,width=15.3cm]{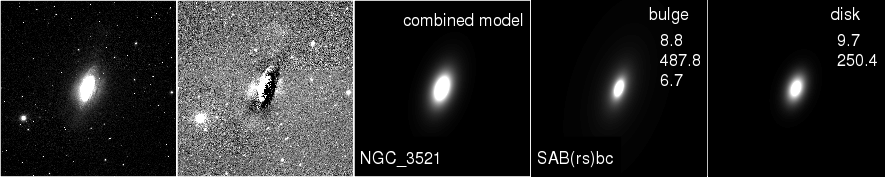}
\includegraphics[height=3.2cm,width=15.3cm]{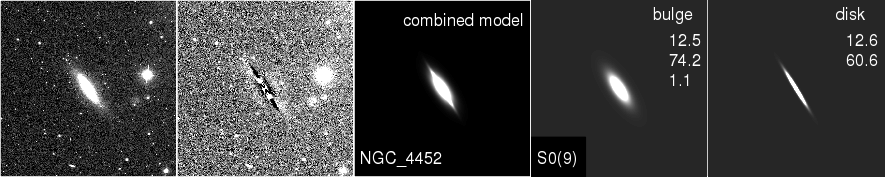}
\includegraphics[height=3.2cm,width=15.3cm]{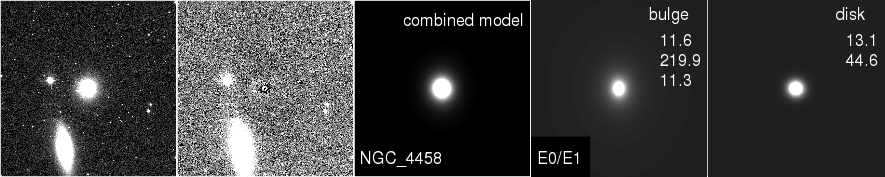}
\includegraphics[height=3.2cm,width=15.3cm]{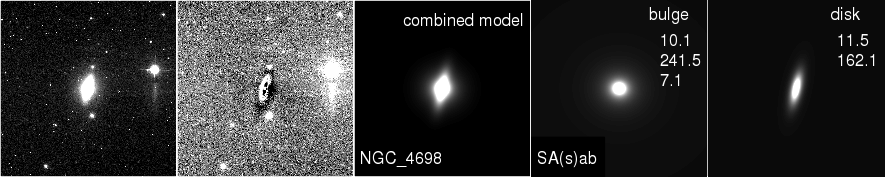}
\includegraphics[height=3.2cm,width=15.3cm]{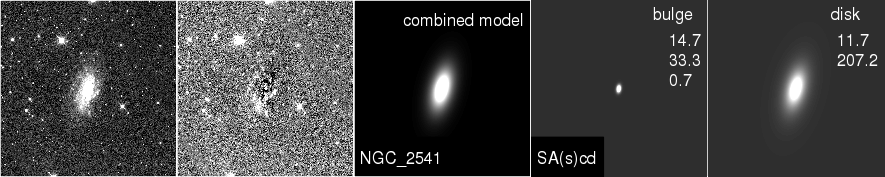}
\includegraphics[height=3.2cm,width=15.3cm]{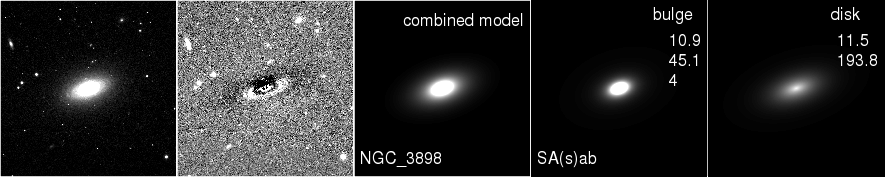}
\includegraphics[height=3.2cm,width=15.3cm]{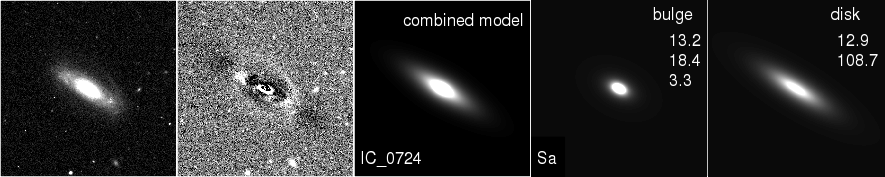}
\caption{A selection of  $r-$band images of the galaxies discussed in Appendix A and some galaxies with good results.  The first column on the left shows the original images, the second column shows  the residuals from the \fitc bulge plus disc fit. 
The third, fourth and fifth columns display the combined model, bulge model (\sersic function) and disc model (exponential function), respectively.  The legends in the fourth and fifth columns show the bulge and disc magnitude, effective radius (in pixels), and bulge \sersic index.}
\label{fig:apend}
\end{figure*}

\section{Notes on Individual galaxies}
\label{sec:notes}

Here, we briefly describe some galaxies with peculiar results.   In Figure~\ref{fig:apend} we present the images and models for some of the following galaxies in order to show the visual appearance of the galaxies that return peculiar results. In the same Figure we have also added three additional galaxies with what we consider good result in order to support a comparison.

\begin{description}

\item[NGC3521] [SAB(rs)bc] A high-inclination galaxy with strong dust features that make it extremely difficult to fit a two-component model. Our two-component model measures a large bulge with $r_{\rm e,b}/r_{\rm e,d}= 1.94$ and $n_{\rm b}=6.7$.  This may not be realistic, but it has been retained in the analysis.
\\
\item[NGC3642] [SA(r)bc] A late-type galaxy for which we measure $r_{\rm e,b}/r_{\rm e,d}= 2.5$ and $n_{\rm b} = 6.7$. However, by examining the equivalent of Fig.~\ref{fig:gal1} for NGC3642, we noticed that that both $r_{\rm e,b}/r_{\rm e,d}$ and $n_{\rm b}$ decrease with redshift until $z=0.03$. After that, the values remain constant until $z=0.12$. The recovered values for this range of redshifts are $r_{\rm e,b}/r_{\rm e,d}= 0.2$ and $n_{\rm b} = 1.5$, which seem much more reasonable. These dramatic changes in parameters could be due to the detailed substructure that is visible in the high-resolution images. As the galaxy becomes more distant these substructure are less pronounced and, as a result, the fit parameters better reflect the properties of the bulge and disc. 
\\
\item[NGC3893] [SAB(rs)c] A similar case to NGC3642. 
\\
\item[NGC4123] [SB(r)c] A late-type galaxy for which our two-component model measures $n_{\rm b} = 10.5$. However, if we add a PSF into the model, the bulge \sersic index is reduced to $1.86$ and $r_{\rm b}$ decreases by $50$\%. The corresponding change in disc effective radius is $\sim 2$\%.  We therefore use the model including the PSF component.
\\
\item[NGC4452] [S0(9)] An almost perfectly edge-on galaxy, which contains a very thin disc (for which our model measures $b/a = 0.09$). The bulge component fits a second elongated component ($b/a = 0.37$) with $n_{\rm b} = 1.08$ and effective radius $20$\% larger than the thin disc.  The properties of the second component are more consistent with a thick disc, rather than a bulge. However, in the spirit of avoiding specal cases, we retain both components in the analysis.
\\
\item[NGC4458] [E] An elliptical galaxy with an internal structure. As a result, the exponential component fits the inner part of the galaxy and the free-\sersic component measures $n_{\rm b} = 11$.  Due to the peculiar fitting results, we attempted to fit this galaxy with various initial parameter values. However, the final results always remained the same, and agree for both \fitb and \fitc methods.
\\
\item[NGC4459] [S0]  After various attempts with different initial values, we failed to fit a significant second component for this lenticular galaxy. In analysis we therefore only include the bulge component. 
\\
\item[NGC4698] [SA(s)ab] A spiral galaxy with peculiar structure: the bulge is elongated perpendicular to the main disc. We measure structural parameters of $r_{\rm e,b}/r_{\rm e,d}= 1.5$ and $n_{\rm b} = 7.1$, consistently using both the \fitb and \fitc methods, and the redshifted images.  However, these are not in agreement with other studies which focus on the unique structure of this galaxy.
\\

\item[UGC08041] [SB(s)d] is a similar case to NGC4123. The two component model measures $n_{\rm b}=9.37$, while the inclusion of a PSF component reduces this to $n_{\rm b}=1.17$. Both the effective radius of the bulge and disc component change less than $10$\%. We use the model with the PSF component.

\end{description}

\label{lastpage}

\end{document}